\documentclass[11pt]{article} % use larger type; default would be 10pt

\usepackage[utf8]{inputenc} % set input encoding (not needed with XeLaTeX)
\usepackage{booktabs,siunitx}
\usepackage{hyperref}
\usepackage{authblk}
\usepackage{natbib}
\usepackage{bbm}
\usepackage{placeins}
\usepackage{geometry} % to change the page dimensions
\usepackage{graphicx} % support the \includegraphics command and options
\usepackage{array}
\usepackage{xcolor}
\usepackage{multirow}
\usepackage{booktabs}

\usepackage{rotating}
\usepackage{dcolumn} 
\newcolumntype{d}{D{.}{.}{4}} 
\usepackage{hyperref}
\usepackage{booktabs} % for much better looking tables
\usepackage{array} % for better arrays (eg matrices) in maths
\usepackage{paralist} % very flexible & customisable lists (eg. enumerate/itemize, etc.)
\usepackage{verbatim} % adds environment for commenting out blocks of text & for better verbatim
\usepackage{subfig} % make it possible to include more than one captioned figure/table in a single float
\usepackage{amsmath}
\usepackage{amsthm} 
\usepackage{amssymb}
\usepackage{wasysym}
\usepackage{graphicx}
\usepackage{comment}
\usepackage{color}
\usepackage{multicol}
\usepackage{fancyhdr} % This should be set AFTER setting up the page geometry
\usepackage{sectsty}
\allsectionsfont{\sffamily\mdseries\upshape} % (See the fntguide.pdf for font help)
\usepackage[nottoc,notlof,notlot]{tocbibind} % Put the bibliography in the ToC
\usepackage[titles,subfigure]{tocloft} % Alter the style of the Table of Contents

\usepackage{authblk}
\definecolor{NavyBlue}{RGB}{35,35,142}
\definecolor{RawSienna}{RGB}{199,97,20}

\hypersetup{
  colorlinks,%
  citecolor=NavyBlue,%
  filecolor=NavyBlue,%
  linkcolor=RawSienna,%
  urlcolor=NavyBlue
}

\definecolor{highlight}{HTML}{000000}
\def\RR{\mathbb{R}}

\title{Data-driven modeling of beam loss in the LHC} 

\author[a,1]{Ekaterina Krymova\footnote{coressponding author E-mail: ekaterina.krymova at epfl.ch}}
\author[a]{Guillaume Obozinski} 
\author[c]{Michael Schenk} 
\author[b,c]{Loic Coyle} 
\author[b,c]{Tatiana Pieloni} 
 
\date{}

\affil[a]{Swiss Data Science Center, EPFL \& ETH Zürich, INN, Station 14, 1015 Lausanne, Switzerland} 
\affil[b]{Particle Accelerator Physics Laboratory, Institute of Physics, EPFL, Lausanne} 
\affil[c]{CERN, 1211 Geneva, Switzerland}

\begin{document} 
\maketitle

\begin{abstract}

%%% Leave the Abstract empty if your article does not require one, please see the Summary Table for full details.
 
In the Large Hadron Collider, the beam losses are continuously measured for machine protection. By design, most of the particle losses occur in the collimation system, where the particles with high oscillation amplitudes or large momentum error are scraped from the beams. The level of particle losses typically is optimized manually by changing multiple control parameters, among which are, for example, currents in the focusing and defocusing magnets along the collider. It is generally challenging to model and predict losses based on the control parameters due to various (non-linear) effects in the system, such as electron clouds, resonance effects, etc, and multiple sources of uncertainty. At the same time understanding the influence of control parameters on the losses is extremely important in order to improve the operation and performance, and future design of accelerators. Existing results showed that it is hard to generalize the models \textcolor{highlight}{\cite{coyle2018machine}}, which assume the \textcolor{highlight}{regression model of losses depending on control parameters}, from fills carried out throughout one year to the data of another year. To circumvent this, we propose to use an autoregressive modeling approach, where we take into account not only the observed control parameters but also previous loss values. We use an equivalent Kalman Filter (KF) formulation in order to efficiently estimate models with different lags. %As baseline models, we use the recurrent NN approach, Random Forest. 
\textcolor{highlight}{The results on the data from 2017 show} that the proposed modeling significantly improves the prediction quality for the year 2018 and successfully identifies both local and global trends in the losses. 
 
\textit{Keywords: beam losses, accelerator control, predictive model, ARMAX, Kalman filter} %All article types: you may provide up to 8 keywords; at least 5 are mandatory.
\end{abstract}

\section{Introduction}

Excessively high beam losses in the Large Hadron Collider (LHC) \cite{evans2007large,bruning2012large} can cause a quench in the superconducting magnets which will trigger a beam dump. As a result, valuable time and hence integrated luminosity is lost for the physics experiments while the LHC needs to be refilled with beam. On the other hand, better control of losses guarantees  more efficient operation and higher luminosity, and thus greater discovery potential. During the LHC run, the machine operators may change several parameters of the system, such as currents in quadrupole, sextupole and octupole magnets, in order to maximize the beam intensity and thus minimize the particle loss. A better understanding of the dependence between input parameters and time series of beam losses can help to decrease beam losses and improve the operation of LHC and potential future colliders, such as the Future Circular Collider (FCC) \cite{barletta2014future,benedikt2015optimizing}. Machine learning and statistical methods have been extensively used to analyze the data from accelerators and to improve the operation, to cite a few \cite{arpaia2021machine, li2021novel, koser2022input, schenk2021modeling, edelen2020machine, coyle2021detection}. The response of the losses to instantaneous changes in the control parameters can be well modeled by the standard ML techniques \cite{coyle2018machine} based on the LHC fills within the same year, but the generalization of such approaches to the data of the next year was found to be challenging. The goal of the paper is to improve the understanding of the effect of input parameters on losses and to propose an interpretable model which would be general and robust enough to generalize LHC fills from different years.

The majority of the losses occur in the collimation systems in the beam cleaning areas \textcolor{highlight}{in the Interaction Regions (IR)} IR3 and IR7 of the accelerator, where the losses are recorded by beam loss monitors (BLM) \cite{hermes2015betatron}. The collimators at IR7 remove particles with large transverse oscillation amplitudes, whereas those at IR3 are responsible for removing the particles with a momentum error beyond a chosen threshold. Among the two collimator systems, the most active cleaning happens in IR7, therefore we concentrate on modeling of losses of beam 1 recorded by BLM at IR7 and further refer to it as the ``loss". 

Several additional important characteristics can be measured during the machine run, which cannot be directly controlled, but that contain information about hidden nonlinear processes in the beam.  Among such quantities are: heat loads that are proxy to electron cloud effects \cite{zimmermann2002electron}, horizontal and vertical emittances. The emittances describe the spread of the particles in the phase space and are related to the mean physical dimension of the beam in the machine \cite{edwards2008introduction}. The emittance measurements are carried out in time along the beam and special post-processing is used to estimate the average emittance of the whole beam. During operation of the LHC, electron clouds can appear due to the acceleration of electrons in the beam pipe by the proton bunches causing an avalanche process, which leads to the heating of the beam pipe and magnets, increased emittance and potentially to beam instabilities \cite{zimmermann1997simulation}. 

 Tune variables are related to the frequency of betatronic oscillations in the machine.  Tune distribution for the particles in the beam demonstrates irregular behavior, thus the number of particles that experience resonance is not straightforward to know.  Tunes are corrected through a dedicated feedback system and mainly depend on the strength of the quadrupole magnets, although can be also affected by the quadrupole component in the main dipoles, by the sextupole component from the main dipoles and sextupole corrector magnets \cite{aquilina2015tune}.  
 
 As most of the non-linear effects in the beam physics are indirectly related to the change of input parameters it appears natural to rely on the information contained in the past observation, therefore we exploit the models of Vector AutoRegressive Moving Average with eXogenous input variables (VARMAX) type and also compare them with the models which map the input variables directly to the losses. We will train a common model on the data of multiple time series from one year of different beam filling schemes. By doing that we presume that random effects, which might be present during several fills, could be averaged out and the model could potentially exctract the universal mechanisms of impacting losses common to different fills, filling schemes and even years.

\begin{figure}[h!]
    \centering 
  \includegraphics[width=0.5\textwidth]{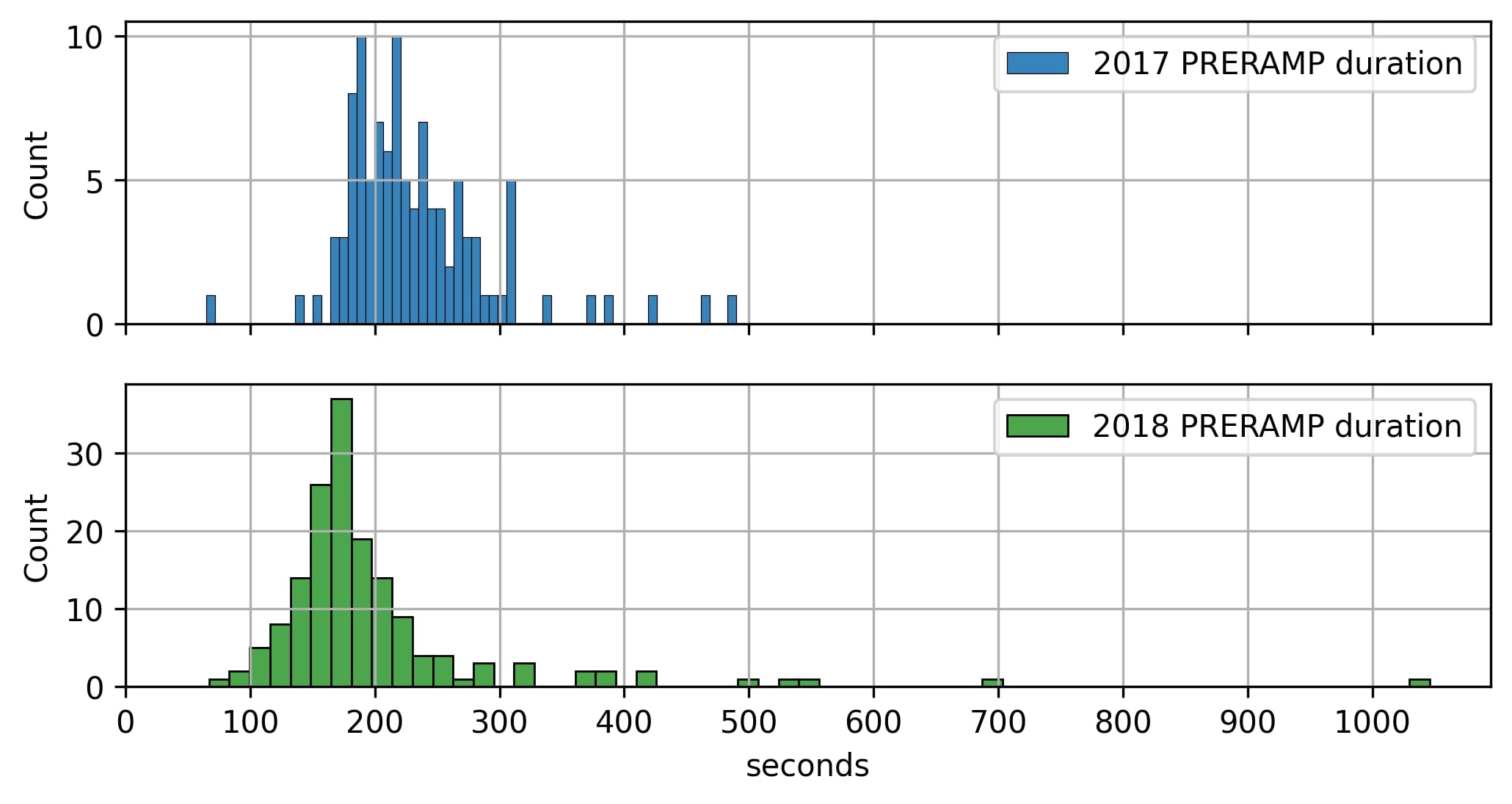}
  \caption{Histogram of PRERAMP time series lengths for the years 2017 and 2018.}
    \label{fig:hist_lengths}
\end{figure}

\textbf{Available data.}  To construct and evaluate the predictive model of losses the \textcolor{highlight}{observations of the losses along with other quantities measured during the} LHC fills of the years 2017 and 2018 are available. Observations are recorded with a frequency 1 Hz.    The fills start with the injection of the beam according to a selected filling scheme,  defining which of the radio-frequency (rf) buckets are filled with protons and which ones are left empty. We have selected the filling schemes which are occurring most frequently among all the encountered schemes in the data as we expected that the properties of the injected beam could vary depending on the type of injection. %For 2017 the selected prevailing schemes are:    {\small{'25ns\_2556b\_2544\_2215\_2332\_144bpi\_20inj'}},
%        {\small{'25ns\_1916b\_1909\_1042\_1560\_112bpi\_20i8b4e'}},
%    and  {\small{'25ns\_1868b\_1866\_1089\_1749\_128bpi\_17i8b4e'}},
%    for 2018: 
%{\small{'25ns\_2556b\_2544\_2215\_2332\_144bpi\_20injV2'}} and 
%       {\small{'25ns\_2556b\_2544\_2215\_2332\_144bpi\_20injV3'}}. 
All together for the selected schemes the data of 106 fills are available in 2017, and 161 fills are recorded for 2018.
We focus on the data for the beam 1 during the ``PRERAMP" mode of the machine fill. For each fill during the years 2017 and 2018 the time series of ``PRERAMP" mode vary in length (see Figure \ref{fig:hist_lengths}).

\begin{figure}[h!]
    \centering
    \includegraphics[width=1\textwidth]{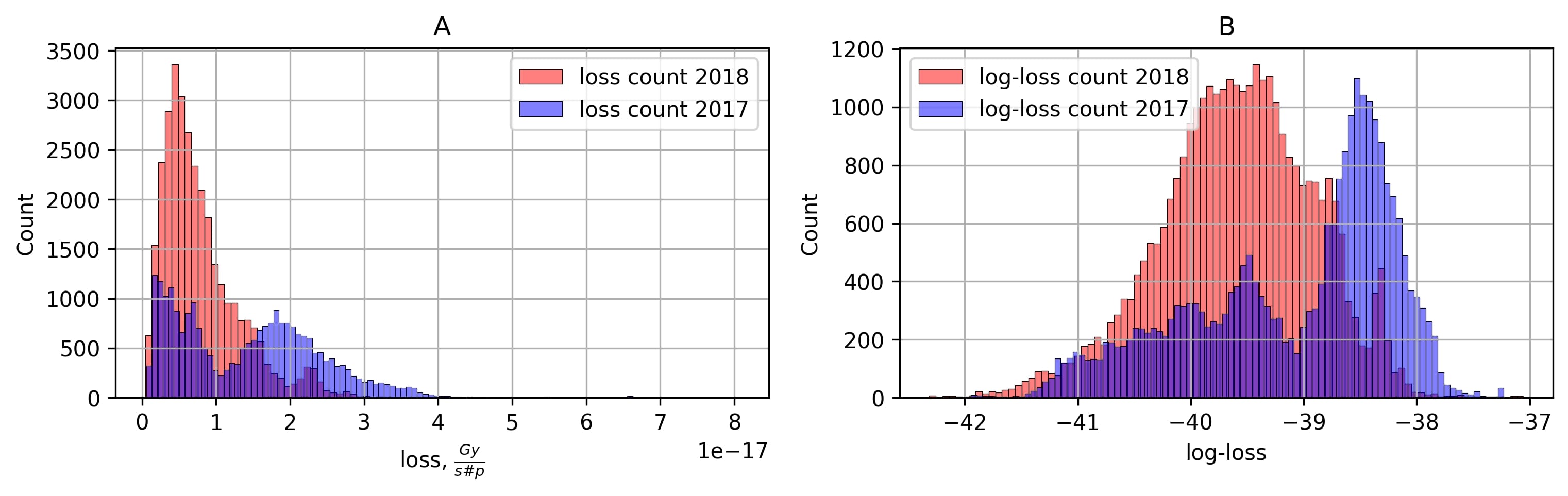} 
     \caption{A: original losses (normalized by intensity),  B: log-losses (normalized by intensity), for the years 2017 and 2018. The log-transform helps to reduce the skewness. Note that several modes on the pictures are due to changes of the losses level because of changes in control parameters, see e.g. Fig. \ref{fig:fill6243}}
    \label{fig:loss_hist}
\end{figure}

Throughout the paper, we assume that logarithmic transformation is applied to losses normalized by intensity. Further, we omit ``log" and ``normalized by intensity" while mentioning losses. Logarithmic transform is generally applied to reduce the skewness of the distribution, see the loss in original scale in Figure \ref{fig:loss_hist}A and the log-transformed loss in Figure \ref{fig:loss_hist}B. For the losses, the log-transformation is partially motivated by the fact that losses normalized by intensity are produced from the particle count data. For the convenience of further analysis, we assume that the log losses are Gaussian. Alternatively, one could follow the Generalized Linear Model approach, assuming the Poisson distribution of the count data, e.g. as in \cite{fokianos2009poisson}.

\begin{figure}[h!]
    \centering
    \includegraphics[width=1\textwidth]{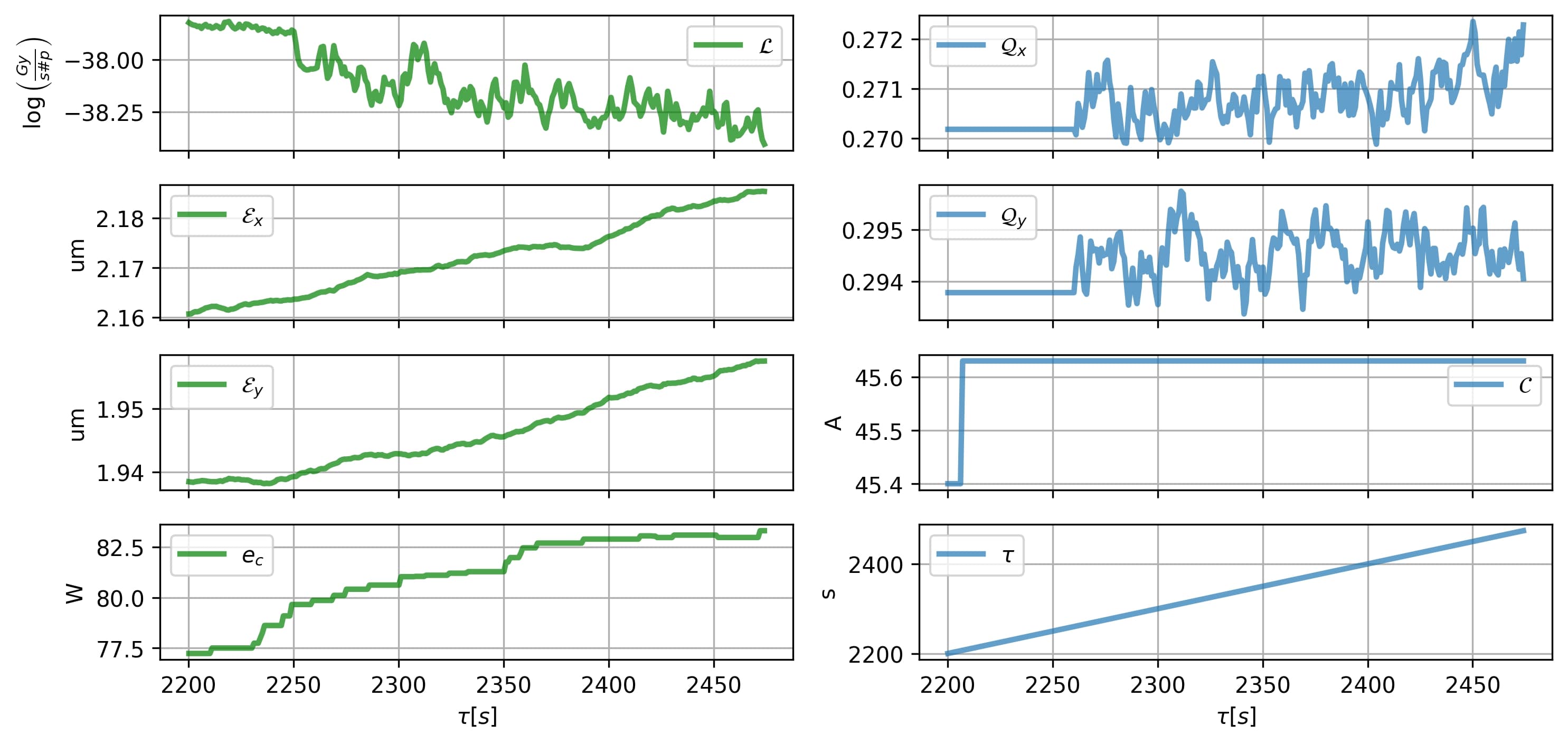} 
     \caption{An example of PRERAMP mode time series for the fill 6243 in 2017. \textcolor{highlight}{In blue time series of vertical and horizontal tunes and octupole currents are shown, observations of the loss, vertical and horizontal emittance and the sum of heatload measurements is shown in green}.}
    \label{fig:fill6243}
\end{figure}

Observations of several quantities are recorded during the experiment, among which we will use the following variables as the input parameters for the loss models 
\begin{itemize}
\item $\mathcal{Q}_x$ and $\mathcal{Q}_y$ ---  vertical and horizontal tunes,%:\\ {\small{'LHC.BOFSU:TUNE\_B1\_H\_fit'}}, {\small{'LHC.BOFSU:TUNE\_B1\_V\_fit'}},
\item $\mathcal{C}$ --- current in octupole magnets. % {\small{'RPMBB.RR13.ROF.A81B1:I\_MEAS'}}.
\end{itemize}
Non-controlled observed variables are 
\begin{itemize}
\item $\mathcal{L}$ --- logarithm of loss divided by intensity at BLM 6L7 for the beam 1,% \\ {\small{'BLMEI.06L7.B1E10\_TCHSH.6L7.B1:LOSS\_RS09'}}; 
\item $\mathcal{E}_x$, $\mathcal{E}_y$ --- horizontal and vertical emittance,%{\small{'EMIT:B1\_H\_MEAN'}},
      %{\small{ 'EMIT:B1\_V\_MEAN'}},
\item  $e_c$ --- sum of the heatload measurements,% {\small{'ecloud'}},
\item $\tau$ --- time since the beam injection.
\end{itemize}
For an example of evolution of the loss and other variables  during the ``PRERAMP" mode within one fill in the year 2017 see Figure \ref{fig:fill6243}. Typically, one significant modification of current and tunes occurs during ``PRERAMP".
 
\begin{figure}[h!]
    \centering
    \includegraphics[width=1\textwidth]{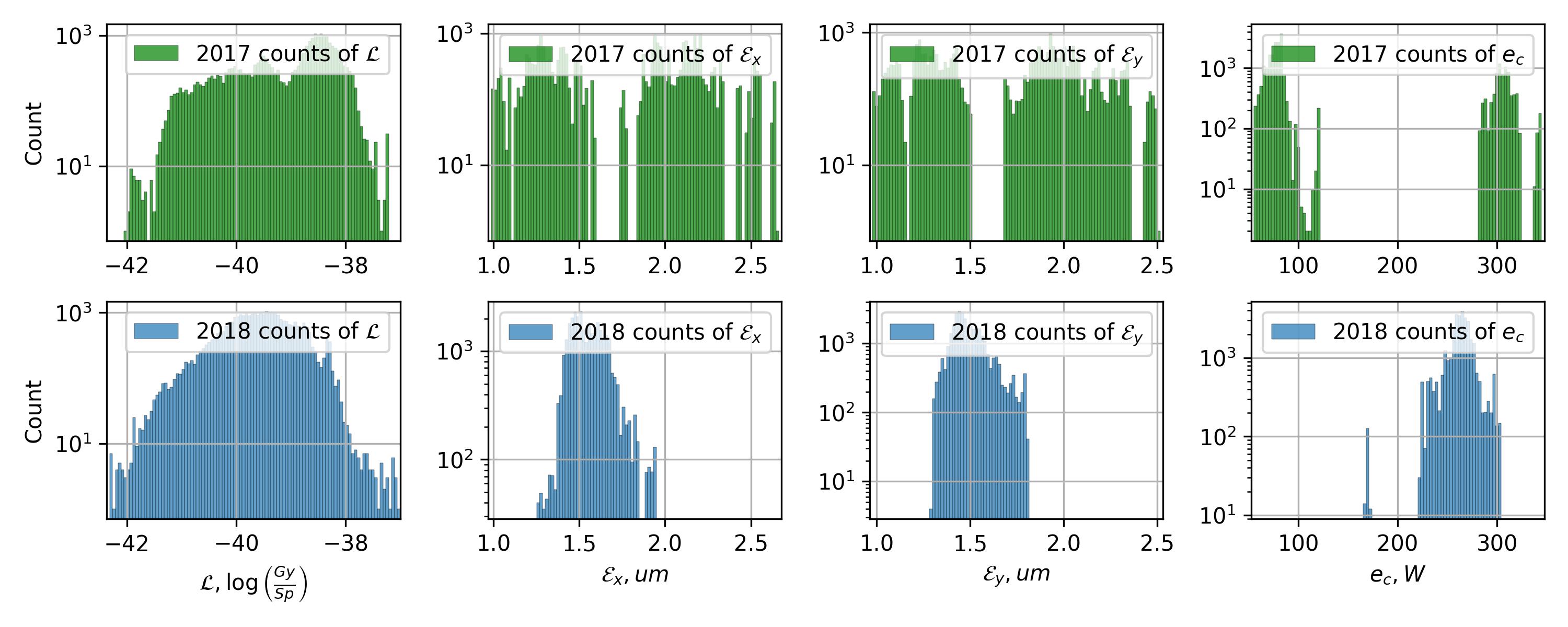}     
\caption{Histograms of the observations in 2017 and 2018. 
A: logarithm of loss normalized by intensity, B: horizontal emittance, C: vertical emittance, D: heat load induced by electron cloud. Almost all the variables demonstrate different ranges of values for both years, e.g. see \textcolor{highlight}{the sum of the heat loads.}}
    \label{fig:data_distribution_output}
\end{figure}

\begin{figure}[h!]
    \centering
    \includegraphics[width=1\textwidth]{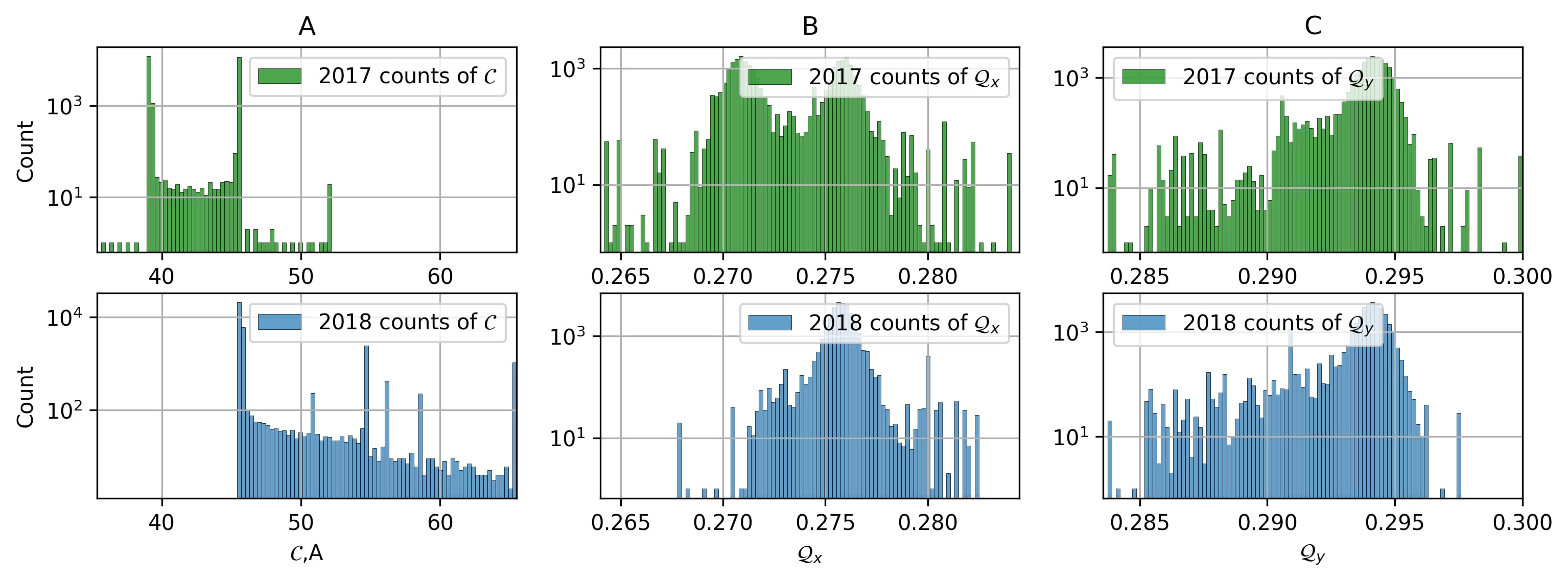}  
    \caption{Histograms of the observations in 2017 and 2018. A: octupole current, B: horizontal tune, C: vertical tune. Almost all the variables demonstrate different ranges of values for both years, e.g. see the octupole current $\mathcal{C}$.}
    \label{fig:data_distribution_input}
\end{figure}

\begin{figure}[h!]
    \centering 
  \includegraphics[width=1\textwidth]{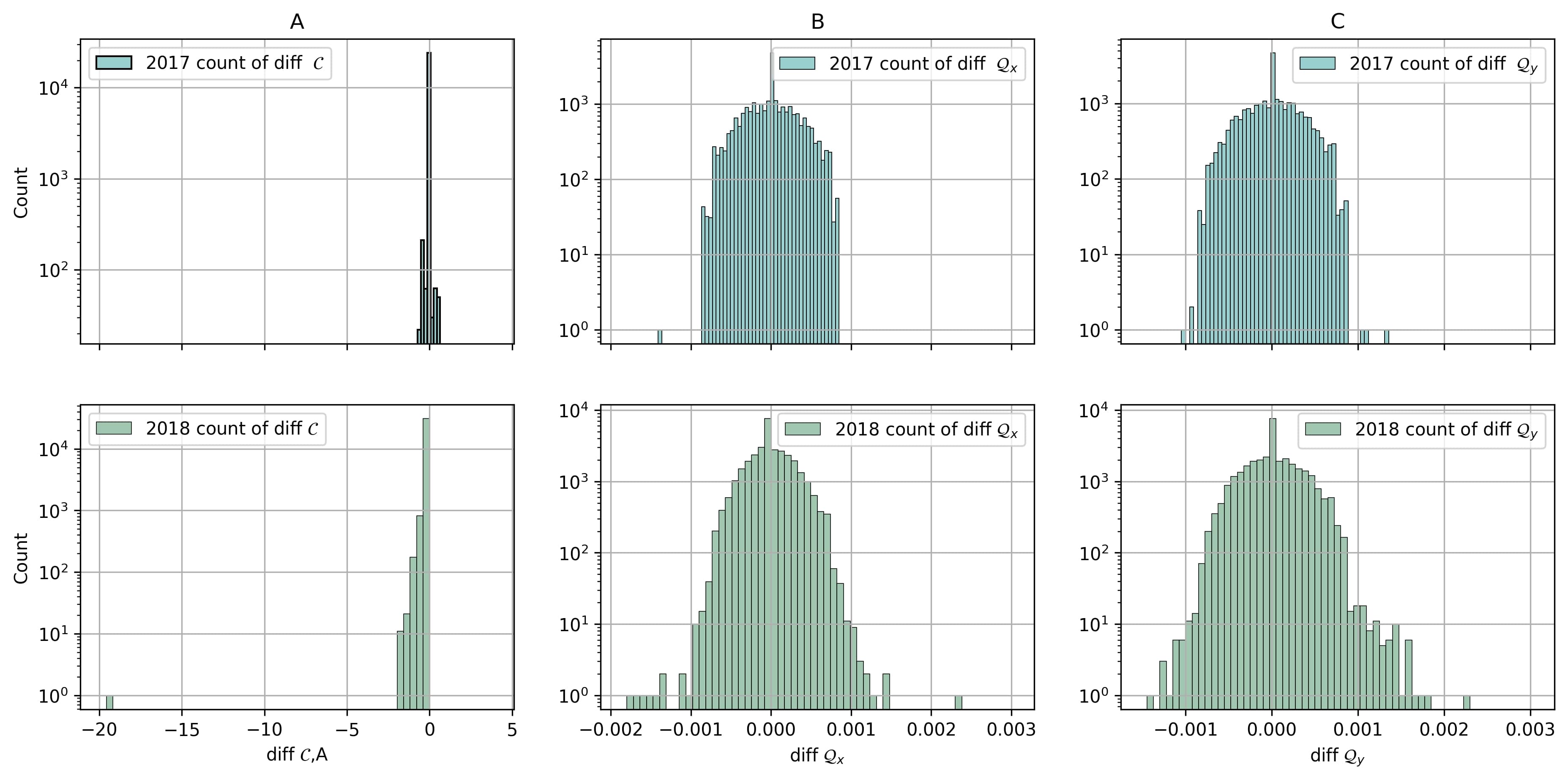}
  \caption{Histogram of increments of  octupole current, horizontal and vertical tunes in 2017 and 2018. The fill 7236 is a big outlier to the left in the histogram of octupole current in 2018.}
    \label{fig:rof_hist}
\end{figure}

The data for the years 2017 and 2018 differ in the ranges of input parameters used and in the level of losses: in Figure \ref{fig:data_distribution_output} non-controlled variables demonstrate different distributions, e.g. losses (Figure \ref{fig:data_distribution_output}A) in 2017 were overall higher than in 2018; input variables, shown in Figure \ref{fig:data_distribution_input}, had different ranges, e.g. see octupole current in Figure \ref{fig:data_distribution_input}A, as well as increments of input variables in Figure \ref{fig:rof_hist}, e.g. octupole current  was only decreasing in 2018, both increasing and decreasing in 2017. Some of the control parameters, such as octupole current, change quite rarely, e.g. in the data of 2017 it changes from one level to another in ``PRERAMP" time series only in half of the fills. This already suggests that applying the model trained on the data from 2017 to the data of 2018 would involve an extrapolation problem, which is known to be extremely challenging. Therefore the use of simple (robust) modeling methods can be anticipated.

Starting from the injection of the beam, multiple effects occur in the system, which cannot be fully described analytically, and have been so far observed in separate experiments, e.g.~changes in the distribution of particle tunes in time and during the change of control parameters and crossing the resonance lines. Given limited information about such events and about long-term dependence between control variables and losses, we consider a simple, but robust linear approach to model the dependence of losses on its prior history and on controlled and possibly uncontrolled variables. An example of such model in 1D case of modeling losses depending on tunes and octupole current would be 
\begin{equation}
\begin{split}
    \mathcal{L}_t  = & \alpha_1 \mathcal{L}_{t-1} + \alpha_2 \mathcal{L}_{t-2} + \dots \\
                & +\beta_1 \mathcal{Q}_{x,t-1} + \beta_2 \mathcal{Q}_{x,t-2} + \dots \\
                & +\gamma_1 \mathcal{Q}_{y,t-1} + \gamma_2 \mathcal{Q}_{y,t-2} + \dots \\ 
                & +\zeta_1 \mathcal{C}_{t-1} + \zeta_2 \mathcal{C}_{t-2} + \dots \\
                & + \text{ noise}.
\end{split}
\label{KF10}
\end{equation} 
With a general form of correlated noise process, this is an instance of an autoregressive model with moving average and exogeneous variables, a so-called ARMAX model \textcolor{highlight}{\cite{hamilton2020time}}.

Given that losses are closely related to emittances and will be affected by electron cloud, and in spite of the fact that these variables cannot be controlled, we will then consider including them among the exogeneous variables. 

Finally, we will consider a multivariate time series model for the losses together with the emittances and electron cloud induced heat loads to try and predict their evolution jointly from controlled variables.
This is is motivated by the fact that we wish to obtain a model that captures the effect of the control variables so that if we rely on the values of emmitances and electron cloud at some time $t-l$ it should itself be predicted from the control variables at prior time points. This kind of multivariate extension is known as a vector ARMAX -- or VARMAX -- model.

To estimate the parameters of this type of model we will exploit the relationship between VARMAX models and the Kalman Filter (KF).

The paper is organized as follows. In Section 2, after introducing more precisely the form of the different models  we discuss VARMAX models and their relation to state space models and in particular to the Kalman filter. Different parameterizations will lead us to consider a general KF with time-varying parameters, including a KF model with parameters depending on the input variables. We will consider different ways of regularizing the coefficients, and discuss a general Expectation-Maximization (EM) procedure for the estimation of parameters.
Section 3 is devoted to the results of numerical experiments and to comparisons of the models. %Finally, we conclude with Section 4.

\section{Models}

In this paper we consider several variants of the model described in \eqref{KF10}. 

\textbf{KF1.} First, we verify whether it is possible to construct a predictive model of the current losses as a function of the recent histories of the losses and of the control variables, as in \ref{KF1}. More precisely, we consider a 1D linear model of losses 
\begin{align}
    \mathcal{L}_t  = & \alpha_1 \mathcal{L}_{t-1} + \alpha_2 \mathcal{L}_{t-2} + \dots \alpha_p \mathcal{L}_{t-p}  \notag\\\notag
                &+\beta_0 \mathcal{Q}_{x,t} + \beta_1 \delta\mathcal{Q}_{x,t} + \beta_2   \delta\mathcal{Q}_{x,t-1} + \dots + \beta_L \delta\mathcal{Q}_{x,t-L+1}  \\\label{KF1}\tag{KF1}
                & +\gamma_0 \mathcal{Q}_{y,t} + \gamma_1 \delta \mathcal{Q}_{y,t} + \dots \gamma_L \delta \mathcal{Q}_{y,t-L+1}\\\notag
                & +\zeta_0 \mathcal{C}_{t} + \zeta_1 \delta\mathcal{C}_{t} + \dots + \zeta_L \delta \mathcal{C}_{t-L+1}+ {\rm noise},\notag
\end{align}
where $\delta$ stands for taking the first order differences, i.e. $\delta \mathcal{Q}_{x,t} = \mathcal{Q}_{x,t}-\mathcal{Q}_{x,t-1}$,  and $p$ and $L$ are the depths of the histories of observations of the outputs and inputs correspondingly, which we include in the model.

\textbf{KF$1^*$} Since we are given additional observations of emittances and electron cloud (heat load sum), we could include them into the input variables to see whether their presence help to model the losses better, thus we will also consider a model  
\begin{align}
    \mathcal{L}_t  = & \alpha_1 \mathcal{L}_{t-1} + \alpha_2 \mathcal{L}_{t-2} + \dots \alpha_p \mathcal{L}_{t-p}  \notag \\ \notag 
    &+\beta_0 \mathcal{Q}_{x,t} + \beta_1 \delta\mathcal{Q}_{x,t} + \beta_2   \delta\mathcal{Q}_{x,t-1} + \dots + \beta_L \delta\mathcal{Q}_{x,t-L+1} \\\notag 
                & +\gamma_0 \mathcal{Q}_{y,t} + \gamma_1 \delta \mathcal{Q}_{y,t} + \dots \gamma_L \delta \mathcal{Q}_{y,t-L+1}\\ \label{KF12} \tag{KF$1^*$}
                & +\zeta_0 \mathcal{C}_{t} + \zeta_1 \delta\mathcal{C}_{t} + \dots + \zeta_L \delta \mathcal{C}_{t-L+1}\\\notag
                &+\eta_0 \mathcal{E}_{x,t} + \eta_1 \delta\mathcal{E}_{x,t} + \eta_2   \delta\mathcal{E}_{x,t-1} + \dots + \eta_L \delta\mathcal{E}_{x,t-L+1}    \\\notag
                & +\theta_0 \mathcal{E}_{y,t} + \theta_1 \delta \mathcal{E}_{y,t} + \dots \theta_L \delta \mathcal{E}_{y,t-L+1}\\\notag
                & +\kappa_0 {e}_{c,t} + \kappa_1 \delta e_{c,t} + \dots + \kappa_L \delta e_{c,t-L+1} + {\rm noise }.\notag
\end{align}

\textbf{KF4.} Next, we could add the additional variables into the output together with the losses 
\begin{align}
 \begin{pmatrix}
\mathcal{L}_t \\
\mathcal{E}_{x,t} \\
\mathcal{E}_{y,t} \\
{e}_{c,t}   
\end{pmatrix} 
= & A_{1}^{\circ} 
\begin{pmatrix}
\mathcal{L}_{t-1} \\
\mathcal{E}_{x,t-1} \\
\mathcal{E}_{y,t-1} \\
{e}_{c,t}   
\end{pmatrix} 
+ \dots + 
A_{p}^{\circ} \begin{pmatrix}
\mathcal{L}_{t-p} \\
\mathcal{E}_{x,t-p} \\
\mathcal{E}_{y,t-p} \\
{e}_{c,t}   
\end{pmatrix}  \label{KF4} \tag{KF4}
\\\notag
& + 
B_{0}^{\circ}  
\begin{pmatrix}
\mathcal{Q}_{x,t}  \\
\mathcal{Q}_{y,t}  \\
\mathcal{C}_{y,t} \\
\tau_{t} 
\end{pmatrix} + 
B_{1}^{\circ}  
\begin{pmatrix}
\delta\mathcal{Q}_{x,t}  \\
\delta\mathcal{Q}_{y,t}  \\
\delta\mathcal{C}_{y,t} 
\end{pmatrix}
+ \dots
B_{L}^{\circ}  
\begin{pmatrix}
\delta\mathcal{Q}_{x,t-L+1}  \\
\delta\mathcal{Q}_{y,t-L+1}  \\
\delta\mathcal{C}_{y,t-L+1} 
\end{pmatrix}
+ {\rm noise}.\notag
\end{align}

\textbf{KF4-quad} is an extension of KF4, and where the matrices $A_i^{\circ}$ and $B_i^{\circ}$ depend on control parameters. In order to fit the model parameters under additional structural assumptions we first consider equivalent formulation of the class of models, which makes it possible to efficiently estimate the parameters. The model is discussed in more details in Section \ref{sec:prox}. 

\subsection{VARMAX and state space modeling}
 
%Starting from the injection of the beam, multiple non-linear effects occur in the system, which cannot be fully described analytically, and have been so far observed in separate experiments, e.g. changes in the distribution of particle tunes in time and during the change of control parameters and crossing the resonance lines.   Given limited information about such events, we consider a simple, but robust linear approach to model the dependence of losses. 

Formally, the models introduced above are all particular instances of a Vector AutoRegressive Moving Average model with eXogenous variables (VARMAX).
VARMAX models can be written as follows:
\begin{equation}
\label{varmax}
    y_t = \sum_{i=1}^{p} A_{i}^{\circ} y_{t-i} + [B_{0}^{\circ} u_{t} +\sum_{i=0}^{L} B_{i+1}^{\circ} \delta  u_{t-i}] +  \sum_{i=0}^{m} C_i^{\circ}\xi_{t-i}, \quad t=1,\dots,T. 
\end{equation} 
The first Vector AutoRegression part represents the belief that the past observations could be predictive of future losses. The second sum, ``X"-part in VARMAX, assumes linear dependence on control variables $u_t$ and their retrospective changes. The last term is a stationary Moving Average process which is a sum of independent random (standard Gaussian) variables (shocks) $\xi_t$ in the past.

A response vector (variable)  $y_t\in\mathbb{R}^{n}$ in VARMAX corresponds to:
\begin{itemize}
    \item a scalar $\mathcal{L}_t$ in the case \eqref{KF1} and \eqref{KF12} and $n=1$,
    \item a vector $[\mathcal{L}_{t},\mathcal{E}_{x,t} ,\mathcal{E}_{y,t-p},{e}_{c,t}]^{\top}$ with the loss, horizontal and vertical emittances and electron cloud ($n=4$) for the case of \eqref{KF4}  and KF4-quad. 
\end{itemize}
We will further assume that $y_t$ is a vector with a 1D case as a sub-case.
The control variable $u_t$ contains different sets of variabes depending on the considered model: 
\begin{itemize}
    \item $u_t \in\mathbb{R}^{q} = [\mathcal{Q}_{x,t},\mathcal{Q}_{y,t}, \mathcal{C}_{y,t}, \tau_t]^{\top}$ a vector with  horizontal and vertical tunes, currents in octupole magnets, and time passed since the end of injection observed at time $t$.  
\item Vectors $\delta u_{t-l}$  contain $l$-lagged differences of $u_{t-l}$, i.e. $\delta u_{t-l} = [\delta\mathcal{Q}_{x,t-l},\delta\mathcal{Q}_{y,t-l},  \delta \mathcal{C}_{y,t-l}]^{\top}$.
\end{itemize}
For estimation we further denote stacked matrices in exogenous term as $B^{\circ} = [B_0^{\circ},B_1^{\circ},\dots,B_L^{\circ}]$ and stacked vector of all exogenous components as $\nu_{L,t} = [u_{t}^{\top},\delta  u_{t}^{\top}, \dots u_{t-L}^{\top}]^{\top}$. 
In these notations \eqref{varmax} reads as 
\begin{equation}
\label{varmax2}
    y_t = \sum_{i=1}^{p} A_{i}^{\circ} y_{t-i} + \sum_{i=1}^{p} B^{\circ}_i \nu_{L,t-i} +  \sum_{i=0}^{m} C_i^{\circ}\xi_{t-i}, \quad t=1,\dots,T. 
\end{equation}
Motivated by VARMAX, we will further consider more general state space models, where the dimension of hidden state could be different from dimension of observations.

\subsection{State space models}
State space models represent the state of a dynamical system by a latent variable, which varies in time and is different from the input and output variables. The most well known model in this family is the Kalman Filter model. State space models are relevant to model time series with rich structure, and there is in particular a well known connection between (V)ARMAX models and Kalman filters that we will exploit in this work.

%\begin{equation}
%    y_t = \sum_{i=1}^{p} a^{\circ} _{i} y_{t-i} +\sum_{i=0}^{m-1} c^{\circ} _i\xi_{t-i}, \quad t=1,\dots,T,
%\label{arma}
%\end{equation}

Consider the 1D autoregressive moving average ARMA(3,2) model (with lag parameters $p=3$, $m=2$):
\begin{equation}
    y_t = a^{\circ} _{1} y_{t-1}+ a^{\circ}_{2} y_{t-2}+ a^{\circ}_{3} y_{t-3} + c^{\circ}_0\xi_{t}+c^{\circ}_1\xi_{t-1}+c^{\circ}_2\xi_{t-2}, \quad t=1,\dots,T,
\label{arma}
\end{equation}
where $\xi_t$ are independent standard Gaussian random variables. 
The ARMA model can be viewed as a special case of the state space model \cite{de2004arma}  with hidden vector $x_t = (x_{1,t}\,x_{2,t}\,x_{3,t})^{\top}$:
\begin{equation}
\begin{split}
y_t = & 
\begin{pmatrix}1& 0 & 0 \end{pmatrix}
\begin{pmatrix}x_{1,t}\\ x_{2,t}\\ x_{3,t} \end{pmatrix}, \\
\begin{pmatrix}x_{1,t}\\ x_{2,t}\\ x_{3,t}\end{pmatrix}  = &  
\begin{pmatrix}
a_1 & 1 & 0 \\
a_2 & 0 & 1 \\ 
a_3 & 0  & 0
\end{pmatrix} 
\begin{pmatrix}x_{1,t-1}\\ x_{2,t-1}\\ x_{3,t-1}  \end{pmatrix} + 
\begin{pmatrix}
c_0 \\
c_1 \\ 
c_2
\end{pmatrix}
\xi_t. 
\end{split}
\label{arma_ss0}
\end{equation}
The first measurement equation describes the relation between observations and hidden state $x_{t}$ of the system, and the second transition equation describes hidden evolution of the state $x_t$. 
The equivalence between \eqref{arma} and \eqref{arma_ss0}  with $a_i= a^{\circ}_i$ and $c_i=c^{\circ}_i$ can be easily seen, if one substitutes $x_{2,t-1}$ and then $x_{3,t-2}$ in the first equation of transition equations using the rest of equations. Thus, in such representation the hidden state components equal lagged output.
The state space representation \eqref{arma_ss0} is not unique, e.g. consider 
\begin{equation}
\begin{split}
y_t = & 
\begin{pmatrix}1& 0 & 0 \end{pmatrix} 
\begin{pmatrix}x_{1,t}\\ x_{2,t}\\ x_{3,t} \end{pmatrix}  + c_0 \xi_{t-1} \\
\begin{pmatrix}x_{1,t}\\ x_{2,t}\\ x_{3,t}\end{pmatrix}  = &  
\begin{pmatrix}
a_1 & 1 & 0 \\
a_2 & 0 & 1 \\ 
a_3 & 0  & 0
\end{pmatrix} 
\begin{pmatrix}x_{1,t-1}\\ x_{2,t-1}\\ x_{3,t-1}\end{pmatrix} + 
\begin{pmatrix}
c_1 \\
c_2 \\ 
c_3
\end{pmatrix}
\xi_{t}. 
\end{split}
\label{arma_ss1}
\end{equation}
In this state space represenation equivalence to \eqref{arma} is slightly less straightforward. One can check that $a_i= a^{\circ}_i$, $c^{\circ}_0= c_0$, $c^{\circ}_1 = c_1-a_1c_0$, $c^{\circ}_2 = c_2-a_2c_0$, $c_3-a_3c_0=0$.

One can see that, for ARMA and corresponding state space representations, each component of the hidden state vector is related to the lagged output, i.e.\ the first component represents the relation to lagged-1 output, and so forth.

In the same way it is possible to write the VARMAX model \eqref{varmax2} in an similar to \label{arma_ss1} state space form:
\begin{equation}
\label{state_space_arma}
\begin{split}
    y_t = D   
    x_{t}+ C_0^{\circ} \xi_t, \\
x_t  = \tilde{A} x_{t-1} +  \tilde{B}\nu_{L,t-i} + \tilde{C}\xi_{t-1}, 
\end{split}
\end{equation}
where $x_t \in \RR^h$ with $h=\max(p,m)$, $\xi_t \in \mathcal{N}(0,I_n)$. 

The matrices in \eqref{state_space_arma} can be defined as follows
$$
D =  \begin{pmatrix} I_h  \\ 0_{n-h\times h}\end{pmatrix}, \text{ if $n>h$, otherwise }  D=\begin{pmatrix} I_n  & 0_{n\times h-n}\end{pmatrix},
$$
where $I_h$ is a squared identity matrix with $h$ columns and rows, $0_{m\times n}$ is a matrix with zeros of the noted size;
 \begin{equation}
\label{state_space_arma0}
\tilde{A}= \begin{pmatrix}
A_1^{\circ} & I_n & 0 &\dots & 0\\
A_2^{\circ} & 0 & I_n &\dots & 0\\
\vdots & \vdots & \vdots &\ddots & \vdots\\
A_{h-1}^{\circ} & 0 & 0 &\dots & I_n\\
A_h^{\circ} & 0 & 0 &\dots & 0 
\end{pmatrix}, \quad
\tilde{B} =  \begin{pmatrix}
B_1^{\circ} \\
B_2^{\circ} \\
\vdots \\
B_{h-1}^{\circ} \\
B_h^{\circ}  
\end{pmatrix}, \quad
\tilde{C} = 
\begin{pmatrix}
C_1^{\circ} \\
C_2^{\circ} \\
\vdots \\
C_{r-1}^{\circ} \\
C_r^{\circ}  
\end{pmatrix} 
  + \begin{pmatrix}
A_1^{\circ} C_0^{\circ}\\
A_2^{\circ} C_0^{\circ}\\
\vdots \\
A_{h-1}^{\circ} C_0^{\circ}\\
A_h^{\circ} C_0^{\circ} 
\end{pmatrix}.
\end{equation}

The state-space representation allows for the use of the efficient inference procedures of the Kalman Filter in the case of Gaussian noise for the known parameters.  When the transition and observation matrices, as well as the noise matrices, are not known, one can use the classical Expectation-Maximization algorithm for their estimation. We discuss briefly their application to the inference and estimation of the parameters in our models.  

\subsubsection{State space model for KF1, KF$1^*$ and KF4}
\label{seq:SSM}

We will be interested in the estimation of the a state space model for the \eqref{KF1},~\eqref{KF12} and \eqref{KF4} models, which will be done in the classical form of the Kalman Filter model: 
\begin{equation}
\begin{split} 
\label{state_space}
y_t &= D x_{t}+ \varepsilon_t, \quad \varepsilon_t\sim \mathcal{N} (0,R),\\
x_t  &= A x_{t-1} + B \nu_{L,t} +  \eta_t, \quad  \eta_t\sim \mathcal{N} (0,V),
\end{split} 
\end{equation} 
where $\nu_{t}$ contains control parameters and their lagged difference up till lag $L$ as in \eqref{varmax}. We assume $x_t \in \mathbb{R}^h$, where $h$ is a multiple of $n$.

We use standard form of the Kalman filter here, instead of \eqref{state_space_arma0} which has a single noise term , as these representations are generally equivalent (see \cite{casals1999fast,casals2012general}). 

\textbf{Remark.}  In \cite{casals2012general}, it was shown that it can be possible to find a matrix $T$ (see the referenced paper for the corresponding algorithm) for the conversion between two state space forms \eqref{state_space} and \eqref{state_space_arma0}, i.e.:  $\tilde{A}= T^{-1}AT$, $\tilde{B} =  T^{-1}B$.
More complicated steps are needed to obtain the matrices ${C}^{\circ}_0$ and  $\tilde{C}$  as \eqref{state_space_arma0} is considered in  a steady-state innovations form (with the same noise vector in both equations).  Namely,  $\tilde{C} =  T^{-1}E \Sigma^{-1/2}$ and ${C}^{\circ}_0=\Sigma^{1/2}$, where  calculation of  $E$ and $\Sigma$ assumes observability condition of the KF model \eqref{eq:conv_KF} and that $P_0-X$ is non-negatively definite (see Theorem 1 in  \cite{casals1999fast}). When these conditions hold, the $h \times n$  matrix $E$  and $n \times n$  matrix $\Sigma$ are defined via $h \times h$ matrix $M$, which is a solution of a Ricatti equation:  
$$\Sigma = DMD^{\top} +R, \quad E = AMD^{\top}\Sigma^{-1},$$
$$M = AMA^{\top} + V - AMD^{\top}(DMD^{\top}+R)^{-1}DMA^{\top}.$$ 

We do not enforce mentioned conditions during estimation of parameters in the KF. After the estimation, we check whether the obtained state space representation satisfies the conditions and then obtain \ref{state_space_arma0} with the VARMAX coefficients.

\subsubsection{State space model KF4-quad with control dependent transitions} 
\label{sec:prox}
In the models we considered so far, and which are motivated initialy by a VARMAX model, the exogeneous variables induce linear shifts in state space via the term $B \nu_{t,L}.$ However, another fairly natural way that the control variable (or control parameters) can enter the model is via the autoregressive coefficients of the VARMAX model or via the transition matrices of the state-space model itself. This motivated us to consider a model which combines both effects: we keep a model of the previous general form, but make the matrices $A$ and $B$ now linearly dependent on $u_t.$ We however limit ourselves to an instantaneous effect. 

We thus consider a Kalman filter model of the form:
\begin{equation}
    \begin{split}
 y_t &= D \,x_{t}+\varepsilon_t, \quad \varepsilon_t\sim \mathcal{N} (0,R),\\
 x_t  &=   {A}(u_t)  \, x_{t-1} + {B}(u_t) \, \nu_{t} +\eta_t,\quad \eta_t\sim \mathcal{N} (0,V).
    \end{split}
    \label{eq:conv_KF}
\end{equation}

with matrices ${A}(u_t)$ and ${B}(u_t)$ now being linear functions of the control variables

\begin{equation}
    {A}(u_t) = A_0+\sum_{j=1}^q A_j u_{tj} \quad \text{and} \quad{B}(u_t) = B_0+\sum_{j=1}^q B_j u_{tj}.
\end{equation}.

This model is now non-linear, and in particular it includes cross-terms of the form $u_t u_{t-j,i}$ and $u_t x_{t-j,i}.$

This formulation has however $q+1$ times more parameters for the state transitions, and regularization becomes necessary to prevent overfitting. Several regularizations would be possible, but given that our model is parameterized by several matrices, we propose to use a matrix regularizer that encourages these matrices to be low-rank (or equal to $0$ which is rank $0$.) More precisely, we propose to use trace norm regularization~\cite{abernethy2009new,hou2013linear}. The trace norm of a matrix (aka nuclear norm), is a matrix norm which is defined as the $\ell_1$-norm of the singular values of the matrix. The trace norm $\|{A}\|_*$ of a matrix $A$ can be equivalently defined by 
$$\|{A}\|_*= {\rm tr}((A^{\top}A)^{1/2}),$$ 
where ${\rm tr}$ denotes the trace of a matrix.
Note that the more classical Tikhonov regularization is here ${\rm tr}(A^{\top}A),$ which is equal to the squared $\ell_2$ norm of the singular values of $A$ and that the $\ell_0$ pseudo-norm of the singular values of a matrix $A$ is the rank of the matrix $A.$ So the trace norm is to rank as the $\ell_1$ norm to $\ell_0$. 
The trace norm is a convex regularizer but induces sparsity in the spectrum of the matrix, in a similar way as the $\ell_1$ norm induces sparsity which means that it becomes low-rank.

In the end, we wish for $A(u)$ and $B(u)$ to be low rank but regularizing directly these matrices with the trace norm leads to an optimization problem which is not so easy to optimize. So, instead, we apply the regularization to all the individual matrices $A_j$ and $B_j.$
We denote 

\begin{equation}
    \Omega((A_j,B_j)_{j=0..q})=\gamma_0\|A_{0}\|_* +\delta_0\|B_{0}\|_*+\gamma \sum_{j=1}^q\|A_{j}\|_*+\delta \sum_{j=1}^q\|B_{j}\|_*.
    \label{eq:omega}
\end{equation}

\subsection{Parameter estimation with the Expectation Maximization algorithm}
\subsubsection{EM for KF1, KF$1^*$ and KF4} 
 \label{EM}
 In order to estimate the parameters $\theta= \{A, B, D, R,V\}$ of the KF  we apply the classical expectation maximization algorithm~\cite{moon1996expectation} to model~\eqref{state_space}. The data consists of pairs $(Y,X) = (y_t^{(j)}, x_t^{(j)})_{t=1,\dots, T_j}$, for $J$ time series of fills with lengths $T_1,\dots,T_J$, where only $Y$ is observed. For simplicity we consider the EM algorithm and omit the superscript with the number of fills, and denote $y^T = \{y_1,\dots,y_T\}$. The extension to multiple fills is straightforward. We assume that the observation and hidden states are Gaussian.
The joint probability density of the complete data $p(Y,X|\theta) = p(x_1) \prod_{t=2}^T p(x_t|x_{t-1}) \prod_{t=1}^T p(y_t|x_{t})$
is given by state space equations (we omitted dependence on $\theta$ in the right hand side).  

To find the maximum likelihood estimate one has to maximize the log-likelihood of the data  $p(Y|\theta) =  \log \int p(Y,X|\theta)p(X|\theta) dX$. To achieve this goal, in the EM algorithm, at each iterations one first computes in the E step a lower bound of log-likelihood of the form 
$$ Q(\theta,\theta_0) =   \int  dX p(X|Y,\theta_0) \log p(Y,X|\theta) \leq \log \int p(Y,X|\theta)p(X|\theta) dX,$$
which is maximized in the M step.

EM alternates between a step that computes a lower bound based on the conditional distribution of the hidden variables given observations with the model parameters determined at the previous iteration, and a  with  maximization step of this lower bound with respect to the parameters of KF, as follows:
\begin{itemize}
\item E step.  To complete the E-step, we have to  estimate the distribution of hidden state given observations  for the fixed $\theta_0$ from the previous step. 
In the Gaussian case, the mean and covariance estimates, as well as  a lag-one-covariance-smoother \cite{welling2010kalman} can be easily calculated via Kalman filtering and smoothing equations. % $\hat{x}_t^T = \mathbf{E}[x_t |y^T]$, covariance $\hat{P}_t^T = \mathbf{E}[(x_t-\hat{x}_t^T)(x_t-\hat{x}_t^T  )^{\top}|y^T]$ and a lag-one-covariance-smoother  $\hat{P}_{t,t-1}^T= \mathbf{E}[(x_t-\hat{x}_t^T )(x_{t-1}-\hat{x}_{t-1}^T)^{\top}|y^T]$. 
The lower bound on the log-likelihood then takes the form
\begin{equation*}
\label{F_estep}
    \begin{split}
        Q(\theta,\theta_0) %& \int  dX p(X|Y, \theta_0) \log p(Y,X|\theta) \\
      = & -\log \det \Sigma_0 - \mathbf{E}_{\theta_0}\left[(x_1-\tau)^{\top} \Sigma_0^{-1}(x_1-\tau)|y^T\right] \\
       -&(T-1)\log \det V -  \mathbf{E}_{\theta_0}\left[\sum_t (x_t-Ax_{t-1}-Bu_t)^{\top}V^{-1}(x_t-Ax_{t-1}-Bu_t)|y^T\right] \\
       -& T\log \det R - \mathbf{E}_{\theta_0}\left[\sum_t (y_t-Dx_t)^{\top}R^{-1}(y_t-Dx_t)|y^T\right],
    \end{split}
\end{equation*}
where $\tau$ and $\Sigma_0$ are the parameters of the Gaussian distribution of the initial hidden state $x_1$.   
\item M step: Maximize  $Q$ with respect to $\theta= \{A, B, D, R,V\}$.
\end{itemize} 
The steps of EM algorithm for the Kalman Filter can be found in the Supplementary Material.

\subsubsection{EM for KF4-quad}

% In order to enforce a trace norm constraint  on a state-transition matrix, note that 
% $$\|A(u)\|_* < \|A_0\|_* + 
% \max_{k}\|u_k\| \sum_{j=1}^{q} \|A_j\|_*.$$
% If we assume that the control variables are standardized such as 
% $\max_{k}\|u_{t,k}\|\leq 1, \text{ for all } t.$, then it is sufficient to suggest that the trace norm of every matrix of the coefficients is bounded:
% $\|A_l\|_*<\tilde{r}/(q+1)$, $l=0,\dots,q.$

For the model KF4-quad, we maximize the \emph{regularized} log-likelihood, using a very similar EM algorithm.
The E-step is exactly the same (only replacing $A$ and $B$ with $A(u_t)$ and $B(u_t)$ respectively). In the M-step we solve the optimization problem
$$\max_{\theta} Q(\theta, \theta_t)-\Omega((A_j,B_j)_{j=0..q}),$$
where $\Omega$ is defined in~\eqref{eq:omega}.
This optimization problem cannot be solved in closed form but it can be solved efficiently using proximal gradient ascent. See Supplementary Materials for details.

% Therefore to impose additional trace norm constraints, we regularize the problem and optimize the parameters by EM algorithm. 
% E-step closely follows the one for KF in Section \ref{EM}  with minor change due to non-constant transition and exogeneous effects matrices. The main modification is required in the M-step of the EM algorithm in Section \ref{EM} as additional regularization terms appear in the lower bound: 
% $$ Q + \gamma_0\|A_{0}\|_* +\delta_0\|B_{0}\|_*+\gamma \sum_{i=1}^p\|A_{i}\|_*+\delta \sum_{i=1}^p\|B_{i}\|_*.$$ 
% For the trace norm regularization the closed-form solution is not available, for the details on the M step see Supplementary Materials. 

Finally, given that the problem is non convex and that the initialization is therefore important, and given that this model differs from the previous one by the introduction of second order terms  \eqref{eq:conv_KF}, we initialize $A_0$, $B_0$  $D$, $V$ and $R$ with the estimates obtained from model~\eqref{state_space}.

\section{Evaluation}
\subsection{Datasets}
The parameters of the model were estimated using ``PRERAMP" observations from one year and then tested on the data of another year.
The data from 2017 contains 105 time series corresponding each to an LHC fill; in 2018, 144 fills are available.
The duration of the ``PRERAMP" phase varies in 2017 from 65 to 490 seconds, whereas in 2018 it varies from 67 to 1046 seconds, with a typical duration which is slightly larger in 2017, see Figure~\ref{fig:hist_lengths}. 

First, we take the data from 2017 as the training dataset and the data of 2018 as the testing dataset. 
After the selection of the hyperparameters (cf Section \ref{hyper_selection}), the model parameters are computed from the full training dataset. 
Then, we validate the trained model on the data from 2018. Next, we repeat the validation for 2018 data as a training set and 2017 as testing to check whether we can also predict the loss of 2017 from 2018.

\begin{figure}[h!]
    \centering 
  \includegraphics[width=1\textwidth]{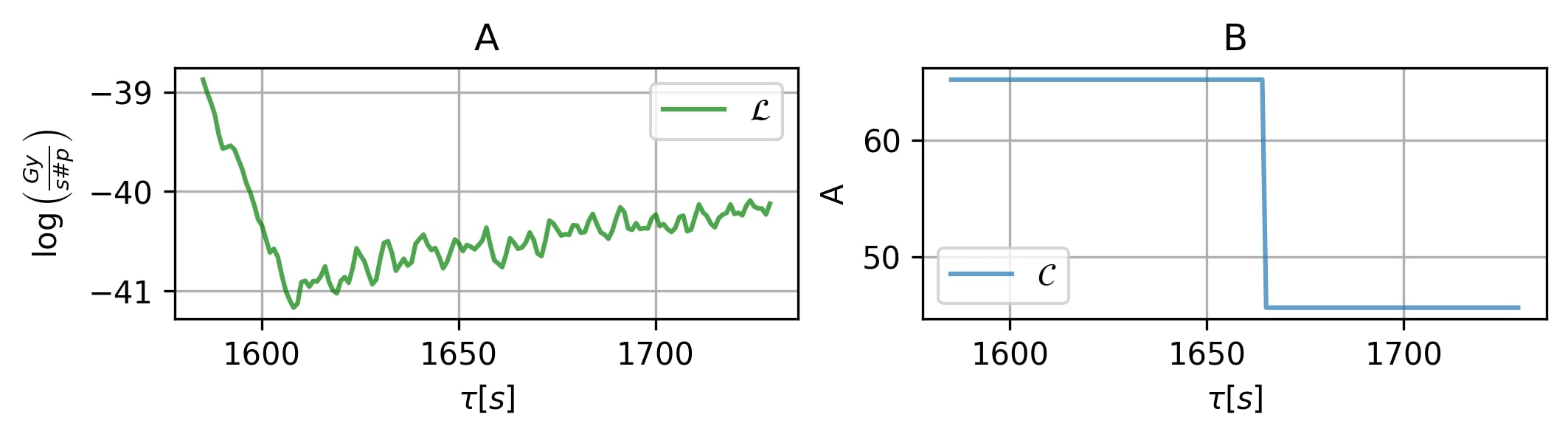}
  \caption{Fill 7236 from 2018, ``PRERAMP" mode, unexpected behavior: no change in losses after large increase in octupole current.}
    \label{fig:fill7236}
\end{figure}

\textbf{Remark}: We excluded fill 7236 from the dataset of 2018 for the second validation. The main reason for exclusion is that during that fill, an abnormally high jump in octupole current occurred (\ref{fig:fill7236}B), which unexpectedly did not lead to a noticeable change in the loss \ref{fig:fill7236}A. There was no other evidence of such events in the datasets and our analysis showed that other variables present in the dataset do not explain such a behavior of the loss. Extending the models for the case of fill 7236 would require additional understanding of the reasons for such loss behavior or more data on similar events. 
This can be additionally seen from Figure \ref{fig:rof_hist}A, where the upper histogram is for the increments of control parameters in 2017 and the lower one is for increments in 2018. 
The change of the octupole current during fill 7236 is close to the value $-20$ A and is distinctly very far from the main range of values. Additionally, we can note that apart from the fill 7236, the changes in octupole current in 2018 are mostly negative, whereas, for the year 2017, the octupole current was both increasing and decreasing. Thus one can anticipate that in terms of octupole input, forecasting the losses of 2018 from the model built from the data from 2017 might be an easier task than when swapping the datasets,  since it involves extrapolation to the larger range of impulse values in the input. 

\textbf{Data Transformations}
For the losses normalized by intensity, we apply a log transformation. Next, the input variables of the training dataset are scaled to be in the interval $[-1,1]$. 
The output variables of the training dataset are centered and normalized. For validation, both input and output are centered and scaled with the parameters obtained for the training dataset.  
\subsubsection{Hyperparameters Estimation}
\label{hyper_selection}
For the KF models, we have two hyperparameters to estimate: the number of lags $L$ in $\nu_t$  in \eqref{state_space} and the dimension of the hidden space $h$. 
To find their estimates we use a 10-fold cross-validation procedure  on the  training dataset to minimize the  mean absolute error (MAE) of the prediction over the parameters in the grid. The MAE for the prediction $\hat{y}_t$ of the ground truth (1D) values $y_t$ is defined as ${\rm MAE} =\frac{1}{T} \sum_t|y_t-\hat{y}_t|$.
We estimate the quality of the predictions of the models built from 9/10 of the fill time series on the rest of the data.
Namely, on each 9/10 of fills, we  an EM estimation of the KF parameters. We set $T_0=10$. Next on the rest of 1/10 of fills, for each fill, we use the KF equations and smoother applied to the first $T_0$-th observation of the time series to get an initial estimate of the hidden process. Starting from $T_0+1$-th observation, we run the KF model state evolution dynamic forward in time to propagate the prediction, giving the control observations as input.   This way, the model ``sees" only the first $T_0$ data points of the output from the fill and the input variable at each new prediction time point. We stack all the predictions for each of the 10 folds to compare them with the corresponding true values, i.e. in each fold we compute MAE: $\frac{1}{\sum_j T_j}\sum_j \sum_{t=T_0+1}^{T_j} |\hat{y}^{(j)}_t -y^{(j)}_t|$ where $T_j$ is a length of the fill $j$, and $j$ runs over the fills in the fold. Finally, the hyperparameters are selected via minimization of the mean MAE across folds.
The hyperparameters selection was carried out on the same intervals of prediction, that is, we fixed the largest history $L_{max}$, and the first data point to predict in the fill of the other year was $L_{max}$ for all the models.
We fixed $L_{max}=90$ to have sufficiently long forecasting horizons and to have enough data to train the models.

The results can be found in Table \ref{agg_rsq}. The hyperparameter selection procedure favored deep histories of the input parameters and their changes, thus the changes in control parameters might have a relatively long-term effect on the loss evolution. 
For KF4-quad,  $h$ and $L$ were set to be the same as the ones found for KF4, and optimization of regularization hyperparameters was carried out in the same way by optimizing the MAE on the grid.
 
\textbf{Remark}:  We compute MAE over different forecast horizons, as opposed to instantaneous one-step ahead forecasting for hyperparameter selection. This is motivated by the fact that minimization of one-step-ahead prediction error tends to select models which better follow local trends. For example, a simple forecast which is just repeating the previous loss observation would often have quite a low one-step ahead forecasting error, whereas for long-term forecasting this is not the case.

\subsection{Evaluation of predictive ability for different time horizons}

We compare the variants of the Kalman Filters: \eqref{KF1}, \eqref{KF12}, \eqref{KF4}, and KF4-quad. 
As evaluation metric of losses prediction for different time horizons  we compute $R^2$-score, which is defined as 
$$R^2 = 1 - \frac{\sum_t(y_t-\hat{y}_t)^2}{\sum_t(y_t-\bar{y}_t)^2},$$
where $\hat{y}_t$ is the predicted value of $y_t$ and $\bar{y}_t=\frac{1}{T} \sum_t {y}_t$ for the models trained on one of the datasets either of 2017 or 2018. 
First, we fit hyperparameters and parameters of the models on the training dataset. For each of the training and testing datasets, for each fill, we fix the horizon $H$. Next, for each time point $t$ of the fill where  $t \in \{T_0+1,\dots,H-t\}$, we use KF equations and smoother to obtain an estimate of the hidden state at $t$, from the preceding $T_0$ observations. Starting from $t$,  we propagate the model to predict the evolution till $t+H$. Thus we get a collection of predictions at horizon $H$ based on the data of different fills. From all the predictions at horizon $H$ we compute a  bootstrap  estimate \textcolor{highlight}{ of the mean $R^2$ \cite{ohtani2000bootstrapping}} obtained from $10^3$ subsamples of $10^3$ predictions and corresponding observations. %The values of $R^2$ close to $1$ signify that the model was able to predict the observation well.

We limited the predictions to the time horizon of $200$ seconds for the dataset of 2017 and to the horizon of $300$ seconds for the dataset 2018, so that 1) for each horizon, the prediction of at least $10$ different fills should contribute to computation of $R^2$ and 2) a number of aggregated prediction was not less than $10^3$.
 
\begin{figure}[h!]
    \centering 
  \includegraphics[width=1\textwidth]{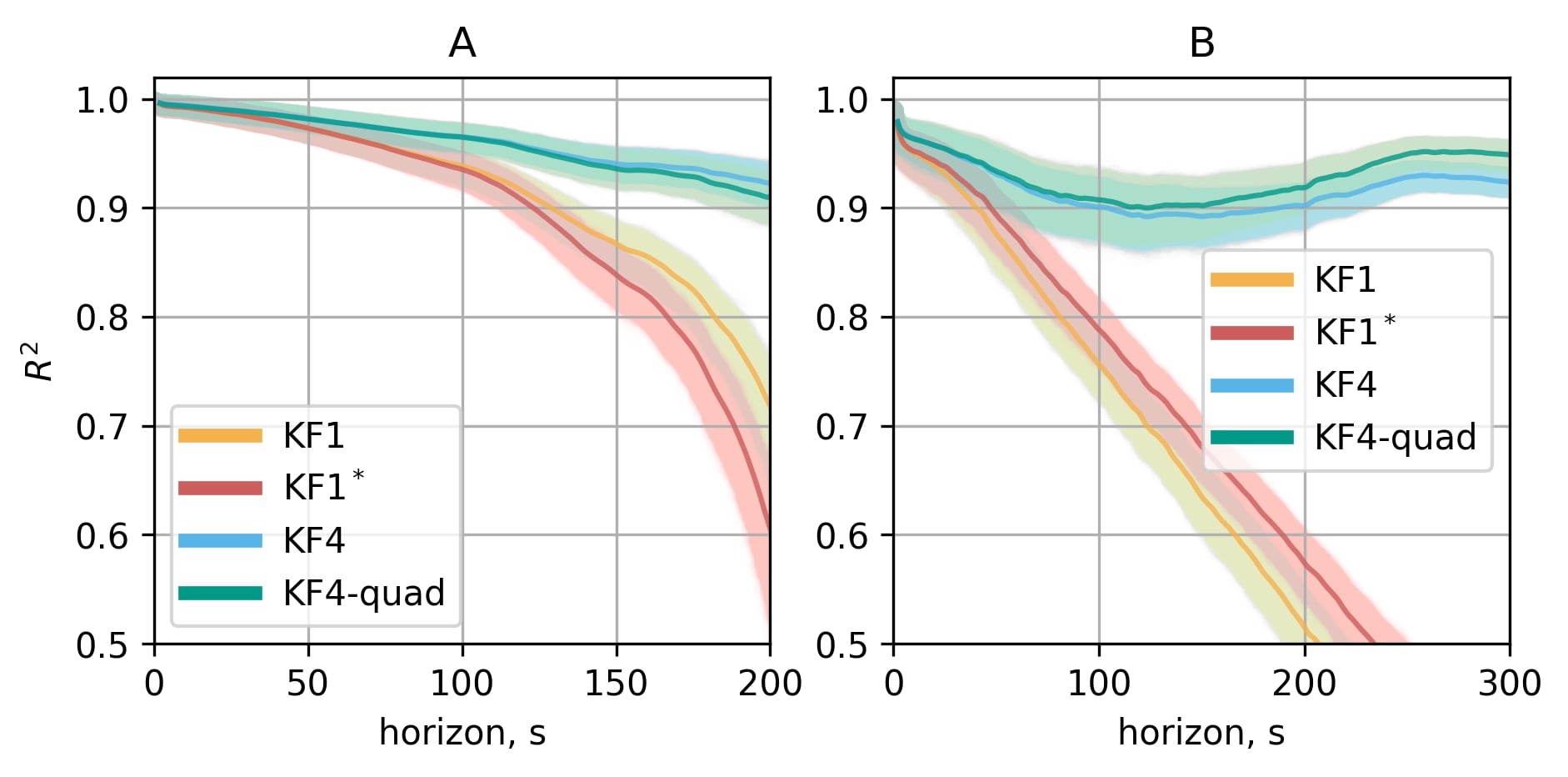}
  \caption{$R^2$-score as a function of the forecast horizon. A:on the training dataset, here from 2017 B: on the test data from 2018. The mean value of $R^2$ is shown for each horizon in the darker color. The mean value was conputer from 1000 bootstrapped estimates of  $R^2$, which are shown in the light color. }
    \label{fig:validation2017}
\end{figure}

\textbf{Training on 2017.} Figure \ref{fig:validation2017} shows how the $R^2$ varies for different horizons $H$ for the models trained on the data from 2017, where \ref{fig:validation2017}A outlines the quality of forecasts on the training dataset, and \ref{fig:validation2017}B shows the quality of forecasts on the testing dataset of 2018.  For the models \eqref{KF1},\eqref{KF12}  with 1D outputs, the results show that they were capable of predicting losses of the other year only on short horizons. Inclusion of additional non-controlling observations, such as emittances and electron cloud in \eqref{KF12} helps to slightly improve the predictive ability in 1D output case on the testing dataset with a certain drop in quality on the training dataset.
Models with additional output components  KF4 and KF4-quad demonstrate  significantly improved performance compared to 1D output models. 
For quite long horizons of the forecast, for both KF4 and KF4-quad  $R^2$ remains high. It is worth noting that the hyperparameters were learned from the dataset of 2017, whereas in 2018  several fills had much longer in time ``PRERAMP" intervals than in 2017 (see Figure\ref{fig:hist_lengths}). Nevertheless,  the propagated KF4 model kept on predicting well on longer horizons. This suggests that overall the model to some extent captures the global trend and its dependence on the input. The bump in $R^2$ for the higher horizons in  \ref{fig:validation2017}B probably could be explained by, first of all, too few fills participating in the estimate, and secondly, most of the changes of input parameters occur on the time horizons smaller than $200$ seconds.  The model KF4-quad demonstrates slightly better performance than KF4 on longer horizons. From the  \ref{fig:validation2017}A the additional regularization helped to reduce the overfitting on the training dataset and improve the $R^2$ on the testing dataset. 

\begin{figure}[h!]
    \centering  
   \includegraphics[width=1\textwidth]{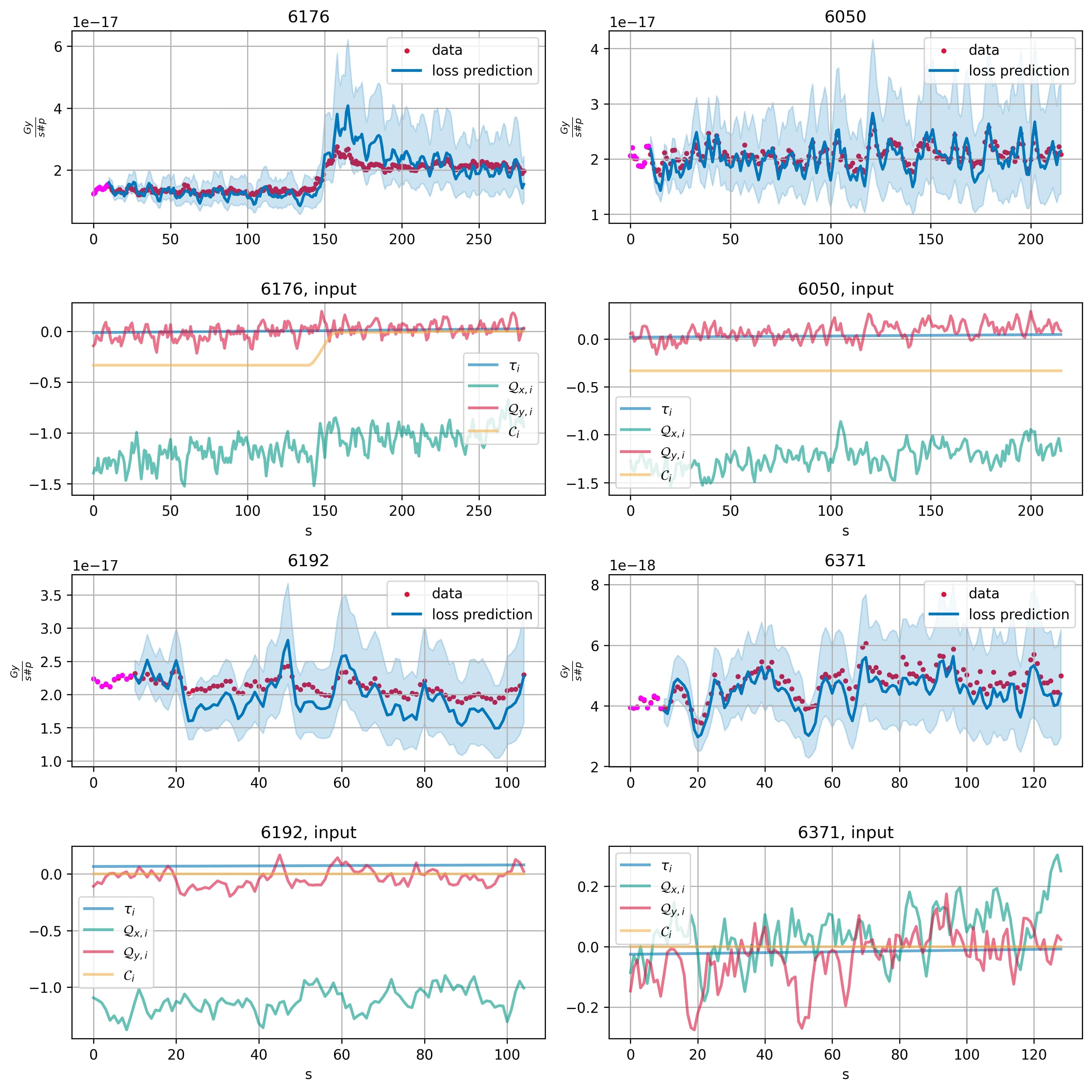}  
  \caption{\label{fig:train2017}KF4 trained on the data of 2017, prediction on the fills of 2018 and corresponding input control parameters. Pink points correspond to $T_0$ observations which the model uses to get initial KF smoother results. Further model propagates without seeing the loss and other output values, with control parameters given as the input. Two standard deviation confidence bands are shown in light blue.} 
\end{figure}

Predictions for selected fills 6672, 6674, 6677, and 6681 of the testing dataset of the year 2018 are shown in Figure \ref{fig:train2017} for the model \eqref{KF4}.  The values of input parameters in the training dataset lie in the interval $[-1,1]$. In Figure \ref{fig:train2017} one can see that the values of some of the input parameters for the testing dataset of 2018 which were standardized to the scale of the training dataset fall outside of interval $[-1,1]$. For the fills 6672, 6674, and 6681 the scaled octupole current decreases from almost $2.5$ to $1$. It is visible that for these fills the model captured the dependence of the loss on the current correctly even outside of the range of the values given during training.

\begin{figure}[h!]
    \centering 
  \includegraphics[width=1\textwidth]{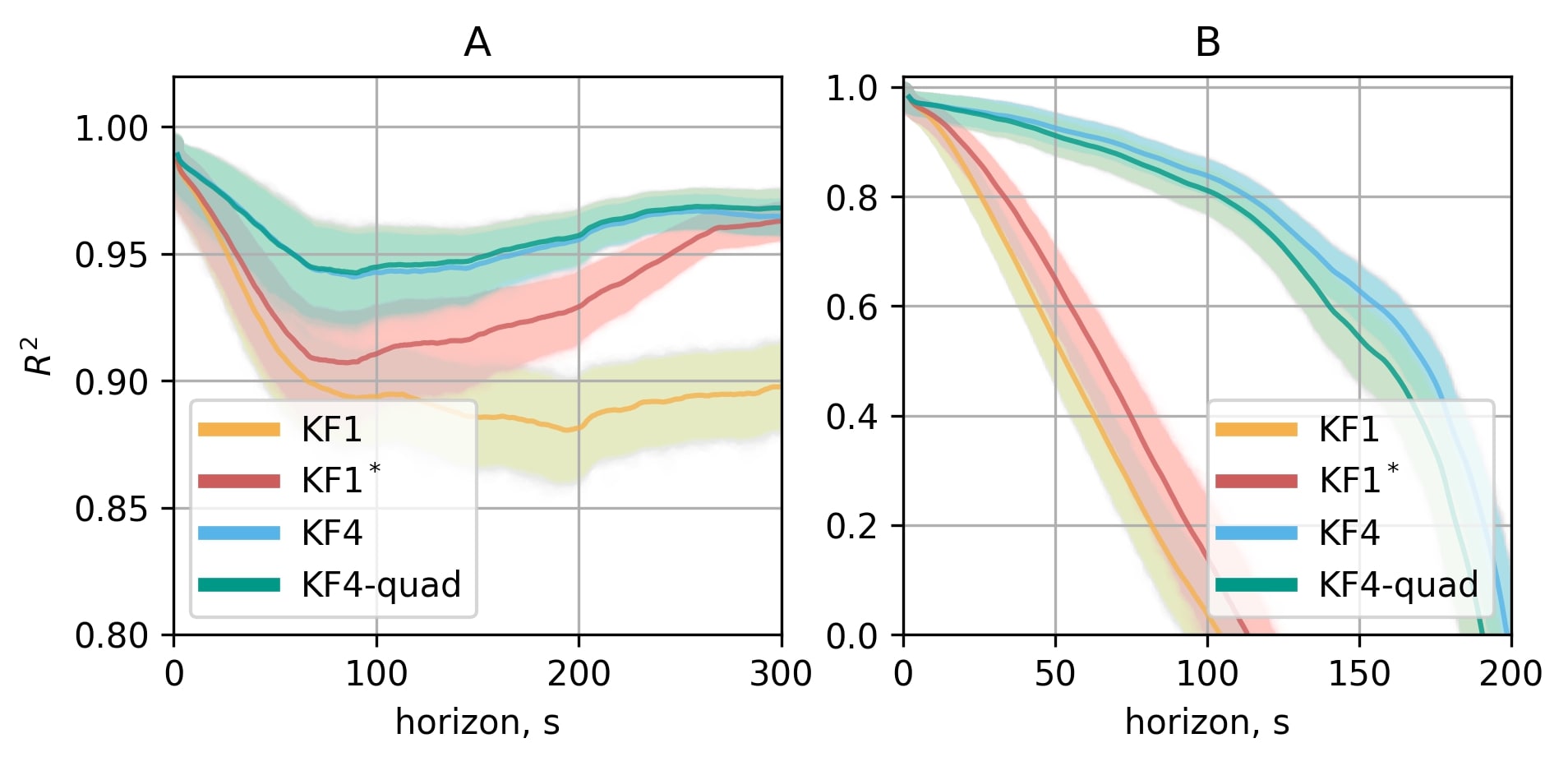}
  \caption{$R^2$-score as a function of the forecast horizon. A: on the training dataset, here from 2018 B: on the test data in 2017.The mean value of $R^2$ is shown for each horizon in the darker color. The mean value was conputer from 1000 bootstrapped estimates of  $R^2$, which are shown in the light color.}
    \label{fig:validation2018}
\end{figure}

\begin{figure}[h!]
    \centering  
   \includegraphics[width=1\textwidth]{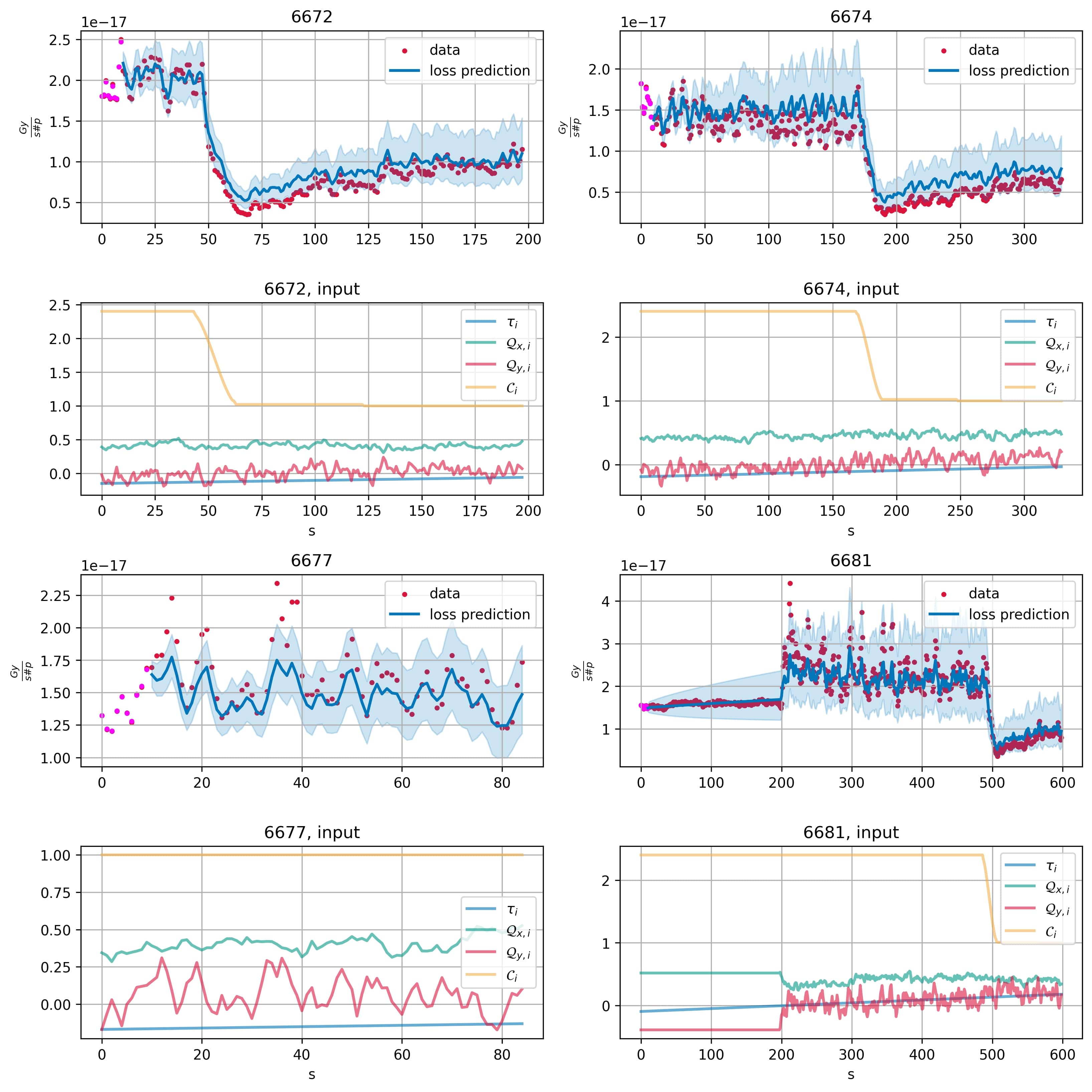}     
  \caption{\label{fig:train2018} KF4 trained on the data of 2018, prediction on the fills of 2017 and corresponding input control parameters. Pink points correspond to $T_0$ observations which the model uses to get initial KF smoother results. Further model propagates without seeing the loss and other output values, with control parameters given as the input. Two standard deviation confidence bands are shown in light blue.}
\end{figure}

\textbf{Training on 2018.} For the models estimated from the data of 2018, the results are shown in Figure \ref{fig:validation2018}. Remind, that the control parameters and their increments in the data of two years have different ranges. The results show that the case of modeling of the loss in 2017 based on the data of the year 2018 is more challenging for the proposed approach. Nevertheless KF4 and KF4-quad show significantly better predictive performance than \eqref{KF1},\eqref{KF12}. Selection of hyperparameters by optimizing MAE in CV for KF4-quad did not lead to improvements compared to KF4. 
Predictions for the fills 6176, 6050, 6192, and 6371 of the year 2017 are shown in Figure \ref{fig:train2018} for KF4 model that was trained on the data of 2018.

\subsubsection{Fitted models} 
\textbf{KF4}. Hyperparameter selection procedure from the data of 2017 led to the \ref{KF4} with the  dimension of hidden process equal $16$, which corresponds to the lag order $4$  in the autoregressive part  and MA parts of the VARMAX model. The selected input parameters increments  history length was $80$.

\begin{figure}[h!]
 \centering 
  \includegraphics[width=1\textwidth]{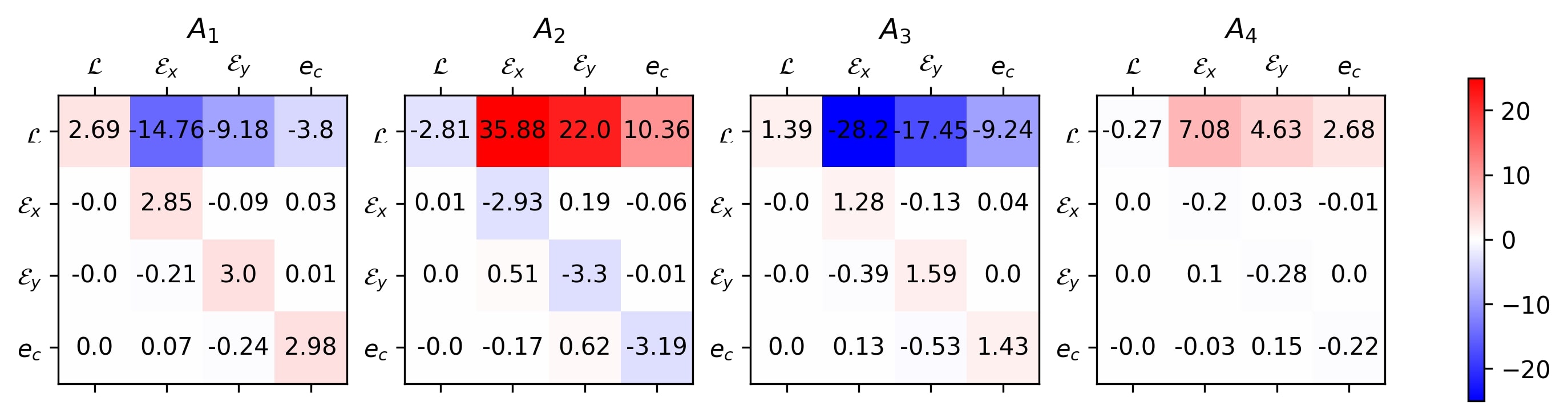}   
   
\caption{\label{fig:ar_coeff} Estimated autoregression matrices in \ref{KF4}.}  
\end{figure}

\begin{figure}[h!]
 \centering 
  \includegraphics[width=1\textwidth]{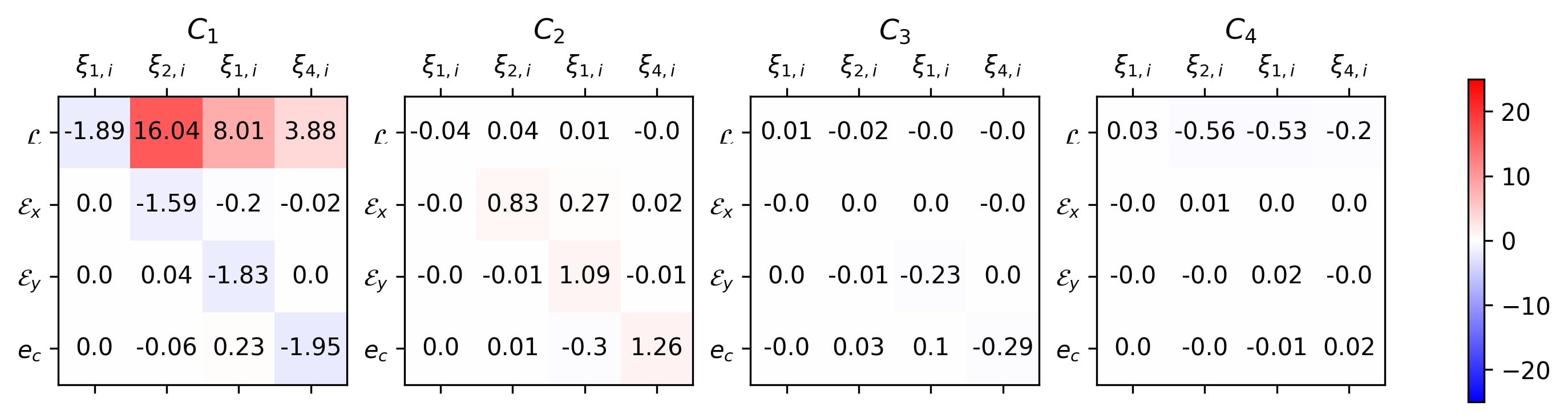}   
\caption{\label{fig:ma_coeff}Estimated moving average part in \ref{KF4}.}   
\end{figure}

After checking that observability condition and condition on the initial value distribution (see remark in Section \ref{seq:SSM}) for the estimated KF4 were satisfied,  we could transform KF4 in the form of \eqref{state_space} to \eqref{state_space_arma0} to obtain the coefficients of equivalent VARMAX model formulation \eqref{varmax2}. Matrices with autoregressive coefficients are shown in Figure \ref{fig:ar_coeff}. One can see that for the losses $\mathcal{L}$, all output variables, including emittances and electron cloud induced heatloads, participate in AR terms. The opposite is not true, in the trained model, all the rest of output values have small coefficients corresponding to the lagged loss variables.  Moving average coefficients \ref{fig:ma_coeff} show that the first lag shocks have the most impact on the losses. 
The loss component of the output shares the shock of "0" lag  with the other output variables.  Further, instead of presenting coefficients for input variables and $80$-lagged increments of input variables, we consider the impulse
response function of the model.

\begin{figure}[h!]
    \centering  
   \includegraphics[width=1\textwidth]{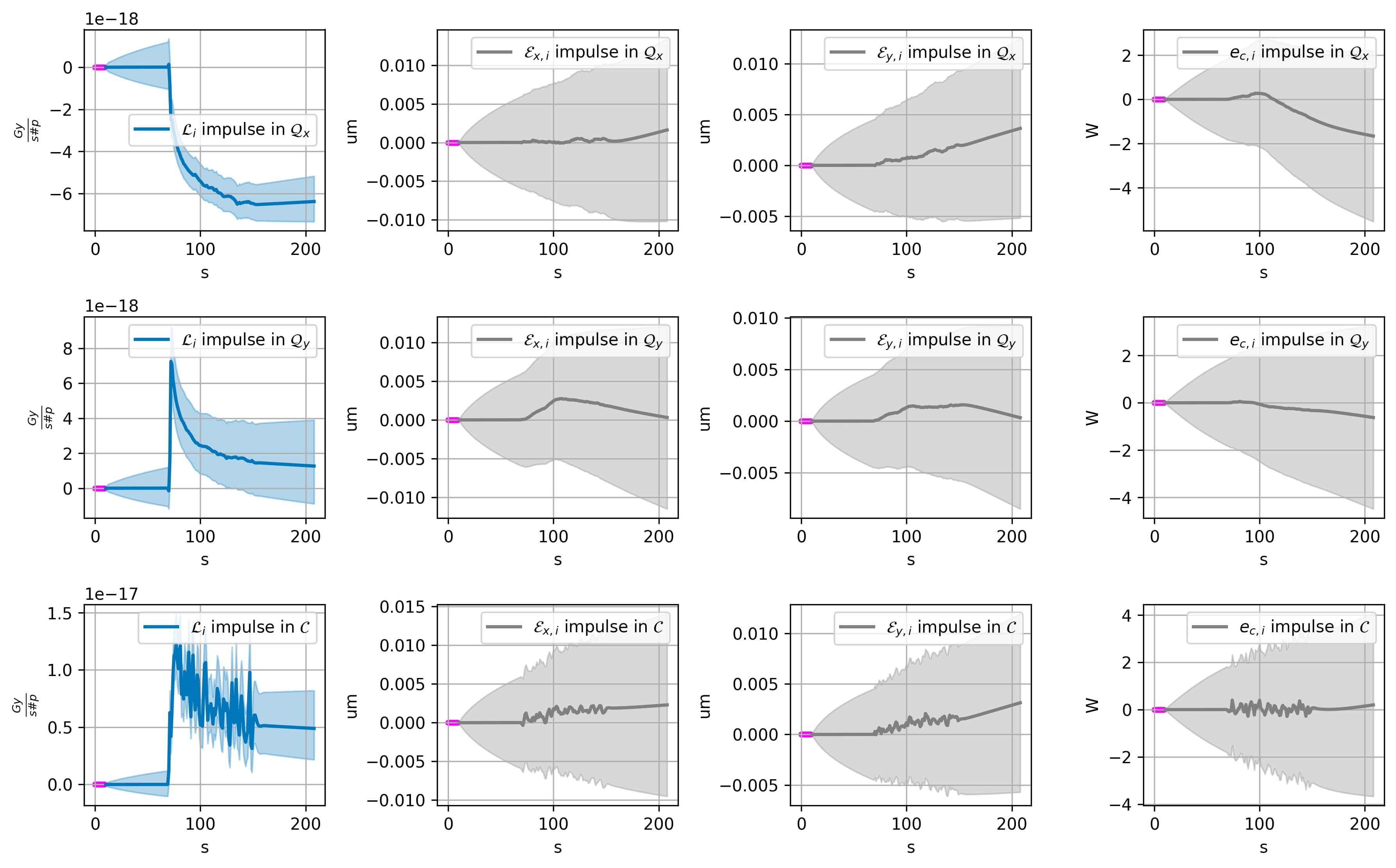}     
    \caption{\label{fig:irf} Impulse response plots for \eqref{KF4} model, impulses in the input parameters: in horizontal and vertical tunes and in the octupole current. The changes in losses and ($2\sigma$-)confidence intervals produced by \eqref{KF4}  are shown in blue colors. The changes in the rest of the output components (horizontal and vertical emittances and electron cloud induced heatloads) and corresponding ($2\sigma$-)confidence intervals are shown in grey.} 
\end{figure}

 \textbf{Impulse response functions}. To analyze how the model \eqref{KF4} responds to shocks in one of the input variables, it is convenient to compute an Impulse Response Function (IRF) \cite{hamilton2020time,belomestny2021bayesian}.
 Figure \ref{fig:irf} demonstrates IRF for \eqref{KF4} trained on the data of 2017: the plot shows the change in output parameters after we modify the input parameters within the ranges in the training dataset. All the values of control and output values were set to median values based on the data from 2017, such that after standardizing them, their values equal zero. The impulse for each of the  variables was taken as $0.5$ of the range of the observations in 2017 at 50 seconds in the Figure \ref{fig:irf}: for horizontal tune $\mathcal{Q}_x$ the impulse was 0.005 from the level 0.272, 
for vertical tune $\mathcal{Q}_y$ the impulse was 0.004 from 0.294, for octupole current the increase was $3.26$A from $39.11$A. After the value of the control variable increases, the model \eqref{KF4} continues to propagate until the outputs stabilize at a certain level. For emittances and electron cloud, IRFs demonstrate that the increase in the octupole current and tunes brings a more steady growth or decrease in values.

\section{Conclusion}
In this paper, we considered a VARMAX approach to model losses in the LHC based on the data of two years for ``PRERAMP" mode. This approach suggests regressing the output not only on the input parameters but also on the previous output observations and previous random effects (shocks). 
Using the relationship with state-space models, we estimated an equivalent Kalman Filter models for VARMAX models with loss as 1D output and with vector-output with loss, horizontal and vertical emittances and electron cloud induced aggregated heat load as components. Additionally, we proposed an extension of the linear KF for the transition matrix and exogenous coefficients dependent on the input variables. 

A hyperparameter selection procedure for lag order parameters in autoregressive and input parts of the models led to a quite long history of input parameters and to the inclusion of a few previous output observations in the autoregression part needed to ensure the small prediction error on a long horizon of forecast. 
The loss model with additional output components fitted on the data of 2017 demonstrated decent performance in predicting the loss measured at IR7 in 2018 for the horizon up to 5 minutes. The impulse response analysis of the estimated model makes it possible to investigate different scenarios of the changes in the control parameters and to understand the effect on the loss predictions learned from the data.

\section*{Conflict of Interest Statement}
%All financial, commercial or other relationships that might be perceived by the academic community as representing a potential conflict of interest must be disclosed. If no such relationship exists, authors will be asked to confirm the following statement: 

The authors declare that the research was conducted in the absence of any commercial or financial relationships that could be construed as a potential conflict of interest.

\section*{Author Contributions}
EK, GO, MS, LC and TP contributed to conception and design of the study. EK performed the statistical analysis and  wrote the first draft of the manuscript. All authors contributed to manuscript revision, read, and approved the submitted version.

\section*{Funding}
This work is funded by the SDSC project C18-07.

%\section*{Acknowledgments}

\section*{Data Availability Statement}
The original contributions presented in the study are included in the article/Supplementary Material; further inquiries can be directed to the corresponding authors.
% Please see the availability of data guidelines for more information, at https://www.frontiersin.org/about/author-guidelines#AvailabilityofData

\bibliographystyle{Frontiers-Vancouver} %  Many Frontiers journals use the Harvard referencing system (Author-date), to find the style and resources for the journal you are submitting to: https://zendesk.frontiersin.org/hc/en-us/articles/360017860337-Frontiers-Reference-Styles-by-Journal. For Humanities and Social Sciences articles please include page numbers in the in-text citations 
\bibliography{reference_list}

\begin{thebibliography}{28}
\expandafter\ifx\csname natexlab\endcsname\relax\def\natexlab#1{#1}\fi
\expandafter\ifx\csname urlstyle\endcsname\relax
  \expandafter\ifx\csname doi\endcsname\relax
  \def\doi#1{doi:\discretionary{}{}{}#1}\fi \else
  \expandafter\ifx\csname doi\endcsname\relax
  \def\doi{doi:\discretionary{}{}{}\begingroup \urlstyle{rm}\Url}\fi \fi
\expandafter\ifx\csname selectlanguage\endcsname\relax
  \def\selectlanguage#1{}\fi

\bibitem[{Coyle(2018)}]{coyle2018machine}
Coyle LTD.
\newblock {\em Machine learning applications for hadron colliders: {LHC}
  lifetime optimization\/}.
\newblock Ph.D. thesis, Grenoble, INP (2018).

\bibitem[{Evans(2007)}]{evans2007large}
Evans L.
\newblock The large hadron collider.
\newblock {\em New Journal of Physics\/} {\bf 9} (2007) 335.

\bibitem[{Br{\"u}ning et~al.(2012)Br{\"u}ning, Burkhardt, and
  Myers}]{bruning2012large}
Br{\"u}ning O, Burkhardt H, Myers S.
\newblock The large hadron collider.
\newblock {\em Progress in Particle and Nuclear Physics\/} {\bf 67} (2012)
  705--734.

\bibitem[{Barletta et~al.(2014)Barletta, Battaglia, Klute, Mangano, Prestemon,
  Rossi et~al.}]{barletta2014future}
Barletta W, Battaglia M, Klute M, Mangano M, Prestemon S, Rossi L, et~al.
\newblock Future hadron colliders: From physics perspectives to technology
  r\&d.
\newblock {\em Nuclear Instruments and Methods in Physics Research Section A:
  Accelerators, Spectrometers, Detectors and Associated Equipment\/} {\bf 764}
  (2014) 352--368.

\bibitem[{Benedikt et~al.(2015)Benedikt, Schulte, and
  Zimmermann}]{benedikt2015optimizing}
Benedikt M, Schulte D, Zimmermann F.
\newblock Optimizing integrated luminosity of future hadron colliders.
\newblock {\em Physical Review Special Topics-Accelerators and Beams\/} {\bf
  18} (2015) 101002.

\bibitem[{Arpaia et~al.(2021)Arpaia, Azzopardi, Blanc, Bregliozzi, Buffat,
  Coyle et~al.}]{arpaia2021machine}
Arpaia P, Azzopardi G, Blanc F, Bregliozzi G, Buffat X, Coyle L, et~al.
\newblock Machine learning for beam dynamics studies at the cern large hadron
  collider.
\newblock {\em Nuclear Instruments and Methods in Physics Research Section A:
  Accelerators, Spectrometers, Detectors and Associated Equipment\/} {\bf 985}
  (2021) 164652.

\bibitem[{Li et~al.(2021)Li, Zacharias, Snuverink, Coello~de Portugal,
  Perez-Cruz, Reggiani et~al.}]{li2021novel}
Li S, Zacharias M, Snuverink J, Coello~de Portugal J, Perez-Cruz F, Reggiani D,
  et~al.
\newblock A novel approach for classification and forecasting of time series in
  particle accelerators.
\newblock {\em Information\/} {\bf 12} (2021) 121.

\bibitem[{Koser et~al.(2022)Koser, Waites, Winklehner, Frey, Adelmann, and
  Conrad}]{koser2022input}
Koser D, Waites L, Winklehner D, Frey M, Adelmann A, Conrad J.
\newblock Input beam matching and beam dynamics design optimizations of the
  {IsoDAR RFQ} using statistical and machine learning techniques.
\newblock {\em Frontiers in Physics\/}  (2022) 302.

\bibitem[{Schenk et~al.(2021)Schenk, Coyle, Giovannozzi, Mereghetti, Pieloni,
  Krymova et~al.}]{schenk2021modeling}
Schenk M, Coyle L, Giovannozzi M, Mereghetti A, Pieloni T, Krymova E, et~al.
\newblock Modeling particle stability plots for accelerator optimization using
  adaptive sampling  (2021).

\bibitem[{Edelen et~al.(2020)Edelen, Neveu, Frey, Huber, Mayes, and
  Adelmann}]{edelen2020machine}
Edelen A, Neveu N, Frey M, Huber Y, Mayes C, Adelmann A.
\newblock Machine learning for orders of magnitude speedup in multiobjective
  optimization of particle accelerator systems.
\newblock {\em Physical Review Accelerators and Beams\/} {\bf 23} (2020)
  044601.

\bibitem[{Coyle et~al.(2021)Coyle, Blanc, Pieloni, Schenk, Buffat, Camillocci
  et~al.}]{coyle2021detection}
Coyle L, Blanc F, Pieloni T, Schenk M, Buffat X, Camillocci M, et~al.
\newblock Detection and classification of collective beam behaviour in the
  {LHC}  (2021).

\bibitem[{Hermes et~al.(2015)Hermes, Redaelli, Jowett, and
  Bruce}]{hermes2015betatron}
Hermes PD, Redaelli S, Jowett J, Bruce R.
\newblock Betatron cleaning for heavy ion beams with ir7 dispersion suppressor
  collimators.
\newblock Tech. rep. (2015).

\bibitem[{Zimmermann(2002)}]{zimmermann2002electron}
Zimmermann F.
\newblock Electron cloud effects in the {LHC}  (2002).

\bibitem[{Edwards and Syphers(2008)}]{edwards2008introduction}
Edwards DA, Syphers MJ.
\newblock {\em An introduction to the physics of high energy accelerators\/}
  (John Wiley \& Sons) (2008).

\bibitem[{Zimmermann(1997)}]{zimmermann1997simulation}
Zimmermann F.
\newblock A simulation study of electron-cloud instability and beam-induced
  multipacting in the {LHC}.
\newblock Tech. rep. (1997).

\bibitem[{Aquilina et~al.(2015)Aquilina, Giovannozzi, Lamont, Sammut,
  Steinhagen, Todesco et~al.}]{aquilina2015tune}
Aquilina N, Giovannozzi M, Lamont M, Sammut N, Steinhagen R, Todesco E, et~al.
\newblock Tune variations in the large hadron collider.
\newblock {\em Nuclear Instruments and Methods in Physics Research Section A:
  Accelerators, Spectrometers, Detectors and Associated Equipment\/} {\bf 778}
  (2015) 6--13.

\bibitem[{Fokianos et~al.(2009)Fokianos, Rahbek, and
  Tj{\o}stheim}]{fokianos2009poisson}
Fokianos K, Rahbek A, Tj{\o}stheim D.
\newblock Poisson autoregression.
\newblock {\em Journal of the American Statistical Association\/} {\bf 104}
  (2009) 1430--1439.

\bibitem[{Hamilton(2020)}]{hamilton2020time}
Hamilton JD.
\newblock {\em Time series analysis\/} (Princeton university press) (2020).

\bibitem[{de~Jong and Penzer(2004)}]{de2004arma}
de~Jong P, Penzer J.
\newblock The {ARMA} model in state space form.
\newblock {\em Statistics \& probability letters\/} {\bf 70} (2004) 119--125.

\bibitem[{Casals et~al.(1999)Casals, Sotoca, and Jerez}]{casals1999fast}
Casals J, Sotoca S, Jerez M.
\newblock A fast and stable method to compute the likelihood of time invariant
  state-space models.
\newblock {\em Economics Letters\/} {\bf 65} (1999) 329--337.

\bibitem[{Casals et~al.(2012)Casals, Garc{\'\i}a-Hiernaux, and
  Jerez}]{casals2012general}
Casals J, Garc{\'\i}a-Hiernaux A, Jerez M.
\newblock From general state-space to {VARMAX} models.
\newblock {\em Mathematics and Computers in Simulation\/} {\bf 82} (2012)
  924--936.

\bibitem[{Abernethy et~al.(2009)Abernethy, Bach, Evgeniou, and
  Vert}]{abernethy2009new}
Abernethy J, Bach F, Evgeniou T, Vert JP.
\newblock A new approach to collaborative filtering: Operator estimation with
  spectral regularization.
\newblock {\em Journal of Machine Learning Research\/} {\bf 10} (2009).

\bibitem[{Hou et~al.(2013)Hou, Zhou, So, and Luo}]{hou2013linear}
Hou K, Zhou Z, So AMC, Luo ZQ.
\newblock On the linear convergence of the proximal gradient method for trace
  norm regularization.
\newblock {\em Advances in Neural Information Processing Systems\/} {\bf 26}
  (2013).

\bibitem[{Moon(1996)}]{moon1996expectation}
Moon TK.
\newblock The expectation-maximization algorithm.
\newblock {\em IEEE Signal processing magazine\/} {\bf 13} (1996) 47--60.

\bibitem[{Welling(2010)}]{welling2010kalman}
Welling M.
\newblock The kalman filter.
\newblock {\em Lecture Note\/}  (2010) 92--117.

\bibitem[{Ohtani(2000)}]{ohtani2000bootstrapping}
Ohtani K.
\newblock Bootstrapping r2 and adjusted r2 in regression analysis.
\newblock {\em Economic Modelling\/} {\bf 17} (2000) 473--483.

\bibitem[{Belomestny et~al.(2021)Belomestny, Krymova, and
  Polbin}]{belomestny2021bayesian}
Belomestny D, Krymova E, Polbin A.
\newblock Bayesian tvp-varx models with time invariant long-run multipliers.
\newblock {\em Economic Modelling\/} {\bf 101} (2021) 105531.

\bibitem[{Lee et~al.(2014)Lee, Sun, and Saunders}]{lee2014proximal}
Lee JD, Sun Y, Saunders MA.
\newblock Proximal {Newton}-type methods for minimizing composite functions.
\newblock {\em SIAM Journal on Optimization\/} {\bf 24} (2014) 1420--1443.

\end{thebibliography}

%%% Make sure to upload the bib file along with the tex file and PDF
%%% Please see the test.bib file for some examples of references
 
\section*{Figure captions}
%#1

%#2

%#3

%#4

%#5

%#6

%#7
 
%#8
 
 %#9
 
 %#10

 %#11

 %#12

 %#13

\section*{Tables}

 %#14
 \begin{table}[h!]
    \centering 
\begin{tabular}{||c |  c c  c  c ||}  
 \hline
 &  $L$, 2017 
 &  $h$, 2017 
 &  $L$, 2018 
 &  $h$, 2018 
  \\ [0.5ex] %KF+EM&
 \hline\hline
 {\small{KF1}}  &  90 & 1  & 55  & 1  \\% & 0.914      \\%0.40 & 
 \hline
 {\small{KF1*}} &  85 & 1  & 80  & 1  \\% & 0.807    \\%0.39 &
  \hline 
 {\small{KF4}}  &  80 & 16 & 80  & 16 \\% & 0.914      \\%0.40 & 
 \hline
\end{tabular}
 \caption{Hyperparameters estimates from 10-fold cross validation for KF variants.} 
\label{agg_rsq}
\end{table}

\FloatBarrier

\newpage 
\pagebreak

\textbf{\large Supplementary Materials}
 
%%%%%%%%%% Merge with supplemental materials %%%%%%%%%%
%%%%%%%%%% Prefix a "S" to all equations, figures, tables and reset the counter %%%%%%%%%%
\setcounter{section}{0}
\setcounter{equation}{0}
\setcounter{figure}{0}
\setcounter{table}{0}
\setcounter{page}{1}
\makeatletter
\renewcommand{\theequation}{S\arabic{equation}}
\renewcommand{\thefigure}{S\arabic{figure}}
\renewcommand{\bibnumfmt}[1]{[S#1]}
\renewcommand{\citenumfont}[1]{S#1}
%%%%%%%%%% Prefix a "S" to all equations, figures, tables and reset the counter %%%%%%%%%%

\section{EM algorithm steps}

\subsection{E-step, Kalman Filter}
\label{EM_KF:Estep} 
For the E-step, we need to obtain  estimates  of $\mathbf{E}_{\theta}[x_t |y^T]$, $\mathbf{E}_{\theta}[x_tx_t^{\top} |y^T]$ and $\mathbf{E}_{\theta}[x_tx_{t-1}^{\top} |y^T]$ for the fixed set of parameters ${\theta}=\{A,B,D,R,V\}$

Given a set of observations $y^T$ the goal is to estimate hidden state, distribution of which depends only on $\hat{x}_t^T$ and $P_t^{T}$. Denote as $\hat{x}_t^{t-1} = \mathbf{E}_{\theta}[x_t| y^{t-1}]$ and $P_t^{t-1} = \mathbf{E}_{\theta}[(x_t-\hat{x}_t^T)(x_t-\hat{x}_t^T)^{\top}| y^{t-1}]$ and $\hat{x}_t^{t} = \mathbf{E}_{\theta}[x_t| y^{t}]$ and $P_t^{t}$ the mean and covariance of $x_t$ conditioned on $y^t$. 

KF equations compute the mean and covariance of the hidden process based on the previous observations:
\begin{align*}
    K_t &= P_t^{t-1} D^{\top} (R + D P_t^{t-1} D^{\top})^{-1},\quad & t\leq T,\\
    \hat{x}_t^{t} & = \hat{x}_t^{t-1} + K_t(y_t-D \hat{x}_t^{t-1}),\quad & t\leq T,\\
    P_t^{t} & = (I - K_tD)P_t^{t-1}, \quad & t\leq T,\\
     \hat{x}_{t+1}^{t} & =  A \hat{x}_t^{t} +  B\nu_t, \quad & t<T,\\
     P_{t+1}^{t} & = AP_t^{t}A^{\top} + V, \quad & t<T.
\end{align*}
The smoother  equations refining the estimates of the states based on information of $T\geq t$ are as follows (backwards in time for $t=T,\dots,1$):
\begin{align*}
    L_t &= P_{t-1}^{t-1} A^{\top} (P_{t}^{t-1} )^{-1},\\
    \hat{x}_{t-1}^{T} & = \hat{x}_{t-1}^{t-1} + L_t(\hat{x}_{t}^{T}-\hat{x}_{t}^{t-1}),\\
    P_{t-1}^{T} & = P_{t-1}^{t-1} + L_t(P_{t}^{T}-P_{t}^{t-1})L_t^{\top}.
\end{align*}
with the initialization from KF equations. 

For the parameter estimation step one would need lag-one covariance  (see \cite{welling2010kalman} for details):
$$
P_{t-1,t-2}^{T} = \mathbf{E}_{\theta}[(x_t-\hat{x}_t^T )(x_{t-1}-\hat{x}_{t-1}^T)^{\top}|y^T] = P_{t-1,t-1}^{T} L_{t-2}^{\top} + L_{t-1} (P_{t,t-1}^{T}- A_t P_{t-1}^{t-1})L_{t-2}^{\top}.
$$
From KF equations and KF smoother we get the following estimates for the fixed set of parameters ${\theta}=\{A,B,D,R,V\}$:
\begin{align*}
 \mathbf{E}_{\theta}[x_t |y^T] =& \hat{x_t}^T\\
 W_t = \mathbf{E}_{\theta}[x_tx_t^{\top }|y^T] =& \hat{P}_t^T +  \hat{x_t}^T[\hat{x}_t^T]^{\top},\\
 W_{t,t-1} = \mathbf{E}_{\theta}[x_tx_{t-1}^{\top }|y^T] = & \hat{P}_{t,t-1}^T +  \hat{x}_t^T[\hat{x}_{t-1}^T]^{\top}. 
\end{align*}

\subsection{M step for the constant coefficients in KF}
 Taking derivatives of the expectation of Q with respect to the components, we find the updated parameter $\hat{\theta}= \{\hat{A}, \hat{B}, \hat{C}, \hat{R}\}$:  
$$\frac{\delta [\mathbf{E} F]}{\delta A} = \sum_{t=2}^T V^{-1}( W_{t,t-1} - A W_{t}-B \nu_t[\hat{x}_{t-1}^{T}]^{\top}) = 0,$$
therefore 
$$\hat{A} = \left[ \sum_{t=2}^T (W_{t,t-1} -B \nu_t[\hat{x}_{t-1}^{T}]^{\top}) \right] \left[ \sum_{t=2}^T  W_{t} \right]^{-1}.$$  
The same way  
$$
\hat{B} = \left[ \sum_{t=2}^T (\nu_t[\hat{x}_{t-1}^{T}]^{\top} \hat{A} \hat{x}_{t-1}^{T}\nu_t^{\top}) \right] \left[  \sum_{t=2}^T  \nu_t\nu_t^{\top} \right]^{-1}$$  
and 
$$
\hat{D}= \sum_{t=0}^{T}\left[y_{t}\hat{x}_t^{\top}\right]\left[\sum_{t=0}^{T} y_{t}y_t^{\top}\right]^{-1}.
$$
The noise terms covariance estimates are as follows: 
$$\hat{R} = \mathbf{E}_{\theta} \frac{1}{T} \sum_{t=1}^{T}   (y_ty^{ \top}_t-\hat{D}\hat{x}^{ }_ty^{\top}_t), \quad 
 \hat{V} = \frac{1}{(T-1)}   \mathbf{E}_{\theta} \sum_{t=1}^{T-1} (x_{t}-\hat{A}{x}_{t-1}-\hat{B}\nu_t) (x_{t}-\hat{A}x_{t-1}-\hat{B}\nu_t)^{\top}.$$

\textbf{Remark.} We additionally transform the estimates such that $\hat{V}$ becomes identity to facilitate the ``inversion" of the state space model to VARMAX.

\textbf{Modification of the M-step.} To accomodate the additional regularization with trace norms, at the M-step, we make a step with the Proximal Newton method \cite{lee2014proximal}, since the objective is quadratic in the optimized arguments and Hessian is non-negatively defined. The Newton step is defined as 
\begin{equation*}
    \Delta A_i = -  [\nabla^2 Q]^{-1}\nabla_{A_l} (- Q), \quad i=0,\dots,q.
\end{equation*} 

The proximal step for the Proximal Newton method with the  step size $l$ is defined as 
$${A}^{(k+1)}_l = {A}^{(k)}-l\Delta_k,$$
with $$\Delta_k = {\rm prox}_{{\rm tr},\gamma}(A^{(k)}-\Delta A_i)-{A}^{(k)},$$ 
where the proximal function for the trace norm regularization is defined via SVD decomposition of the argument matrix, i.e. if 
${\rm SVD} (A) = U{\rm diag(s)}W^{T}$, where $s$ is a vector of eigenvalues of $A$, 
$${\rm prox}_{{\rm tr},\gamma}(A) = U {\rm diag}({\rm prox}_{l_1,\gamma}(s))W^{T},$$
where 
${\rm prox}_{l_1,\gamma}(s) =  {\rm sign}(s) (|s|-\gamma) I(|s|>\gamma)$ is the soft thresholding operator with the threshold parameter $\gamma$, which acts on each component of $s$.

To select a step size $l$, one can use a backtracking line search. It proceeds starting from $l=1$, while (expectation before $L_{h}$ is omitted for simplicity)
$$-Q({A}^{(k+1)}_l) > -Q({A}^{(k)})  
+ \alpha l (-\nabla Q \Delta_k+ h ({A}^{(k)}+\Delta_k )-h ({A}^{(k)})). \quad h(A)=\gamma \|A\|_*$$
The step $l$ is decreased by multiplying with a factor $\beta\in (0,1)$, until violation of condition. The parameter $\alpha$ is selected from $(0,0.5)$ \cite{lee2014proximal}.

\end{document}